\documentclass[12pt]{article} 
\pdfoutput=1
\usepackage{amssymb,amsmath}
\usepackage{longtable} 
\usepackage[american]{babel}   %
\usepackage[latin1]{inputenc}  %
\usepackage{xspace}   %
\usepackage[pdftex]{graphicx}
\DeclareGraphicsExtensions{.jpg,.png,.pdf}

\newcommand{\inserthere}[1]{
    \begin{center}
      \fbox{Insert #1 about here}
    \end{center}
  }

\newcommand{\changespacing}[1]{
  \renewcommand{\baselinestretch}{#1} %
  \small\normalsize
}

\RequirePackage{apacite}
\graphicspath{{pics/}}
\newcommand{\EYEXlong}{E(Y)/E(X)}
\newcommand{\EYEX}{\rho}

\newcommand{\estEYEX}{\hat{\rho}}

\newcommand{\meanx}{\overline{x}}
\newcommand{\meany}{\overline{y}}
\newcommand{\Pmeanx}{\overline{X}}
\newcommand{\Pmeany}{\overline{Y}}
\newcommand{\Bmeanx}{\overline{x}^*}
\newcommand{\Bmeany}{\overline{y}^*}

\newcommand{\Psdmeanx}{\sigma_{\Pmeanx}}
\newcommand{\Psdmeany}{\sigma_{\Pmeany}}
\newcommand{\Psdx}{\sigma_{X}}
\newcommand{\Psdy}{\sigma_{Y}}

\newcommand{\Pcvmeanx}{CV_{\Pmeanx}} 
\newcommand{\Pcvmeany}{CV_{\Pmeany}} 
\newcommand{\Pcvx}{CV_{X}}
\newcommand{\Pcvy}{CV_{Y}} 

\newcommand{\vxx}{\hat{\sigma}^2_{\Pmeanx}}
\newcommand{\vyy}{\hat{\sigma}^2_{\Pmeany}}

\newcommand{\vxy}{\hat{\sigma}_{\Pmeanx,\Pmeany}}

\newcommand{\smeany}{\hat{\sigma}_{\Pmeany}}
\newcommand{\srr}{\hat{\sigma}_{\estEYEX}}

\newcommand{\tquantilsgl}{t_{q}}
\newcommand{\tquantil}{t^2_{q}}
\newcommand{\tunbounded}{t^2_{unbounded}}

\newcommand{\cv}{CV\xspace}
\newcommand{\cvs}{CVs\xspace}
\newcommand{\BCa}{$BC_a$\xspace}
\newcommand{\TheStandardCase}{The standard case}
\newcommand{\WhenCanWeUseRegressionMethods}{When can we use regression methods?} 
\newcommand{\WhenCanWeUseIndices}{When can we use indices?} 
\newcommand{\BewareSpuriousCorrelationsAndFaultyRatioStandards}{Beware: Spurious correlations and faulty ratio standards}

\begin{document}
\title{Ratios: \\ A short guide to \\ confidence limits and proper use}
\author{V.H. Franz\thanks{University of Giessen, Giessen, Germany}}
\date{October, 2007} 
\maketitle 

{\parindent0cm 
\vfill 
\vfill 
\markright{Ratios: Confidence limits \& proper use}
RUNNING HEAD: Ratios: Confidence limits \& proper use.\\ 
\vfill 
\vfill 

Correspondence should
be addressed to: \\ 

Volker Franz \\ 
Justus--Liebig--Universität Giessen  \\
FB 06 / Abt. Allgemeine Psychologie  \\
Otto--Behaghel--Strasse 10F \\
35394 Giessen, Germany \\
\parbox[b]{1.5cm}{Phone:} ++49 (0)641 99--26112\\
\parbox[b]{1.5cm}{Fax:}   ++49 (0)641 99--26119 \\
\parbox[b]{1.5cm}{Email:} volker.franz@psychol.uni-giessen.de \\

\vfill
}
\newpage
\section*{Abstract}

\noindent
Researchers often calculate ratios of measured quantities. Specifying
confidence limits for ratios is difficult and the appropriate methods
are often unknown. Appropriate methods are described (Fieller, Taylor,
special bootstrap methods). For the Fieller method a simple
geometrical interpretation is given. Monte Carlo simulations show when
these methods are appropriate and that the most frequently used
methods (index method and zero--variance method) can lead to large
liberal deviations from the desired confidence level. It is discussed
when we can use standard regression or measurement error models and
when we have to resort to specific models for heteroscedastic data.
Finally, an old warning is repeated that we should be aware of the
problems of spurious correlations if we use ratios.

\newpage
\noindent
In a number of situations, researchers are interested in the ratio of
two measured quantities. For example, in bioassay researchers are
interested in the potency of a drug relative to a standard drug
(\citeNP{Finney_78}). Similarly, whenever we calculate ``percentage
change'' or ``relative change'' we calculate a ratio \cite{Miller_86}.
Another example is inverse prediction in regression analysis. Assume
researchers fit the linear model: $y_i = \alpha + \beta x_i +
\epsilon_i$ (with $i=1,\dots,n$) and then want to predict at which
$x_0$ to expect a certain $y_0$ value. This calculates as
$x_0=(y_0-\hat{\alpha})/\hat{\beta}$ which is again a ratio of the
random parameter estimates $\hat{\alpha}$ and $\hat{\beta}$. Inverse
prediction is often used in calibration procedures (cf.
\citeNP{Kendall_II,Miller_86,Buonaccorsi_01}).

Similar situations arise in psychology and in the neurosciences. 
Ratios of measured quantities have been calculated in the
investigation of perceived speed \cite{HammettTB00},
perceived slant \cite{Proffitt_etal_95},
distance perception \cite{vonderEmdeSGBG98,Proffitt_etal_03},
visual discrimination performance \cite{Watson_Robson_81},
human motor control
\cite{CarrierHE94,SerrienNLLBW02,SerrienW01,TurrellLW01},
psychological and biological bases of stress, drug addiction, and
emotion \cite{LeesN99,MaesCBLN00,ThomasBBM01,YamawakiTF04}.

Specifying confidence limits for ratios is a well--know problem in
statistics with a number of unusual properties.  The classic solution
to this problem is called ``Fieller's theorem''
(\citeNP{Fieller_1940}, see also
\citeNP{Fieller_1944,Fieller_54,Read_83,Buonaccorsi_01}) and is
routinely used in a number of areas (e.g., in bioassay and health
economics, cf.  \citeNP{Finney_78,Briggs_etal_02}). Quite
surprisingly, however, this issue seems to be largely unknown in
psychology and the cognitive neurosciences.
For example, none of the above cited studies used Fieller's method.
Most studies unquestioningly used a method which I will call the
``index'' method and which turns out to require very specific
assumptions about the distribution of numerator and denominator of the
ratio. If these assumptions are not met, the method can lead to
confidence limits with much too small coverage. Other studies used
another ad--hoc method (the ``zero--variance'' method), which is even
more problematic.

The index method is closely related to the use of indices which are
determined on a per observation basis and then processed further as if
they were normal observations. Examples are the body mass index (body
weight divided by height squared) or income per capita (total personal
income divided by total population). Indices are quite frequently used
in medicine and in econometrics and have been in the focus of a long
and heated controversy about spurious correlations
\cite{Pearson_1897,Neyman_79,Kronmal_93}, such that some caution is in
order here. I will sketch the main problems and remedies.

Before discussing the details of the different methods, let me
describe the unusual problems posed by ratios. The main problem arises
from the fact that the function $y/x$ has a singularity at $x=0$.
Therefore, if the denominator is noisy and ``too close'' to zero the
estimate for the ratio goes astray. This problem is so serious that
the probability distribution of the ratio shows unusual behavior. For
example, there neither exists the expected value nor the variance for
the ratio if the denominator is normally distributed. We can only
specify ``pseudo'' values for the expected value and the variance in
cases where the denominator is ``far'' from zero.

A further example for the unusual behavior of ratios is the Cauchy
distribution. This occurs if, in addition to a normally distributed
denominator, the numerator is also normally distributed (and both are
independent and have an expected value of zero). The
probability--density of the Cauchy distribution looks like that of a
normal distribution, but with heavier tails.  Neither the expected
value nor the variance exist for this distribution. Even worse, if we
calculate the mean of independent, identically Cauchy--distributed
variables we find that the mean follows the \emph{same} Cauchy
distribution as each of the individual variables.  That is, the mean
is no more informative than any of the individual values (e.g.,
\citeNP{Johnson_Katz_70}).  This is in strong contrast to the
``typical'' behavior of random variables for which expected value and
variance exist. Typically, calculating the mean of independent,
identically distributed (i.i.d.)  random variables leads to a decrease
of the variance and therefore allows us to use the mean as a better
estimate for the expected value.

Given this unusual behavior, it does not seem surprising that we need
special methods to deal with ratios. I will discuss these methods in
four parts:

The first part (``\TheStandardCase'') discusses confidence limits for
ratios if numerator and denominator are normally distributed. In this
part, I give a simple geometric description of the Fieller method, a
discussion of alternatives to Fieller's method, and of recent
developments in the area of the bootstrap which allows to relax the
assumption of normality. Also, I show in simulations under which
conditions the often used index and zero--variance methods fail and to
which extent this is relevant for the interpretation of existing
studies. For this a number of sample studies are described and the
variability of numerator and denominator in these studies is compared
to the results of the simulations.  (Details about the studies can be
found in the supplementary material provided with this article). A
short summary with recommendations is given at the end of this part.

The second part (``\WhenCanWeUseRegressionMethods'') views ratios as
the special case of a linear model with zero intercept, such that the
ratio corresponds to the slope.  Also, we assume in this part
homoscedastic data.  That is, the variability of the numerator is
assumed to be constant over the range of observations of the
denominator. Linear models allow us to deal with more complex
situations as, for example, the comparison of ratios. I discuss when
we can use standard regression methods and when we have to use the
more complicated measurement error models and show the relationship
between Fieller method and measurement error models.

The third part (``\WhenCanWeUseIndices'') discusses which models are
needed to justify the use of indices and of the index method. We will
see that these models require a special form of heteroscedastic data
with the numerator having larger variability at larger values of the
denominator.

The fourth part
(``\BewareSpuriousCorrelationsAndFaultyRatioStandards'') discusses the
century--old problem of spurious correlations and faulty ratio
standards. Although these problems could appear with any of the
methods discussed in the first three parts of the article, they are
typically discussed in the context of indices. We will see that the
central question is whether we are justified in assuming that the
intercept of a linear model is zero (such that the ratio corresponds
to the slope of the model) or whether we have to assume a non--zero
intercept.

At the end of the article, an overall summary is given which allows
the practitioner to quickly decide which method is appropriate for the
situation at hand.

\section*{\TheStandardCase}
\subsection*{Notation, assumptions, and point--estimate}

Let $X$, $Y$ be random variables with expected values $E(X)$ and
$E(Y)$ and the ratio of interest: $\EYEX := \EYEXlong$. Very often, we
encounter the case of $N$ paired measurements $(x_i,y_i)$ with $i=1
\dots N$ (assumed to be i.i.d.).  When discussing the alternatives to
Fieller's method, we will see that some of these methods are
restricted to paired measurements. For simplicity, I will restrict
most of the discussion to this important case (for generalizations of
the Fieller method to independent samples with unequal variances see
\citeNP{Wu_Jiang_01,Lee_Lin_04}).

Unbiased estimators for the expected values are the sample means
$\meanx$ and $\meany$. Their variances and covariances are estimated
by the usual estimators:
\begin{eqnarray}
\label{eq:var:def}
          \vxx & = & \frac{1}{N} \frac{1}{N-1} 
                     \sum_{i=1}^N (x_i - \meanx)^2 \\
\nonumber \vyy & = & \text{(analogue to $\vxx$)} \\
\nonumber \vxy & = & \frac{1}{N} \frac{1}{N-1} 
                     \sum_{i=1}^N (x_i - \meanx) (y_i - \meany) 
\end{eqnarray}
The coefficients of variation (\cv) for the individual values are:
$\Pcvx := \frac{\Psdx}{E(X)}$ and $\Pcvy := \frac{\Psdy}{E(Y)}$. The
\cvs for the sample means are $\Pcvmeanx := \frac{\Psdmeanx}{E(X)}$ and
$\Pcvmeany := \frac{\Psdmeany}{E(Y)}$.

We assume that $(\Pmeanx,\Pmeany)$ is (approximately or exactly)
distributed as bivariate normal. Note, that due to the central limit
theorem it is often a good approximation to assume the sample means to
be normally distributed even if the individual values are not. For the
bootstrap methods it is possible to relax the assumption of normality,
see the discussion there. For generalizations of the Fieller method to
non--normal distributions, as for example the Gamma, Poisson, or
Weibull distributions, see \citeA{Cox_67} and \citeA{Wu_etal_05}.
An intuitive point--estimate for the ratio of interest is:
\begin{equation}
\label{eq:point:est}
\estEYEX=\frac{\meany}{\meanx}
\end{equation}
This estimator is often used in conjunction with the different methods
to determine confidence limits, as described below (an exception is
the index method). As we are dealing with a ratio, this
estimator shows unusual behavior: First, neither expected value nor
variance exist and the probability distribution is complex (cf.
\citeNP{Marsaglia_65} and \citeNP{Hinkley_69,Hinkley_70}). In cases
where the denominator has a small \cv we can specify ``pseudo'' values
for expected value and variance. We do this by truncating the
distribution such that the denominator cannot get close to zero.
Second, the estimator is biased.  This can be seen by performing a
second order Taylor expansion on the ratio. Certain corrections have
been proposed
\cite{Beale_62,Tin_65,Durbin_59,Rao_81,Dalabehera_Sahoo_95} but in
practical situations they do not seem to perform much better than the
estimator in Equation~(\ref{eq:point:est}) \cite{Miller_86}, such that
they will not be discussed here. Both problems are attenuated with
larger sample sizes because then the \cv of the denominator gets smaller.

\subsection*{Fieller method}

The central statistics of the Fieller method (\citeNP{Fieller_1940},
see also \citeNP{Fieller_1944,Fieller_54,Read_83,Buonaccorsi_01}) can
be derived as follows: Because the difference of normal variables is
also normally
distributed, the term %
$\meany - \EYEX \meanx$ %
is normally distributed. Dividing this term by the appropriate
estimator of the standard deviation gives us the statistics:
\begin{equation}
\label{eq:fieller:T0}
T_0=\frac{\meany - \EYEX \meanx}{\sqrt{\vyy - 2 \EYEX \vxy + \EYEX^2 \vxx}}
\end{equation} 
which follows approximately or exactly a Student--t--distribution with
$df$ degrees of freedom.

In most cases this relationship is only approximate and the
t--distribution corresponds to the normal distribution (with
$df=\infty$). The relationship is exact if the following conditions
are met: (a) $(\Pmeanx,\Pmeany)$ is exactly normally distributed (b)
the variance--covariance matrix is known up to a proportionality
constant: $\sigma^2$ (c) the proportionality constant is estimated by the
estimator $\hat{\sigma}^2$ independent of $(\meanx,\meany)$, such that $\frac{df
  \hat{\sigma}^2}{\sigma^2}$ is distributed as chi--square with $df$ degrees of
freedom. In this case, the t--distribution has $df$ degrees of freedom
(cf. \citeNP{Buonaccorsi_01}).

To obtain confidence limits for $\EYEX$, we calculate the set of
$\EYEX$ values for which the corresponding $T_0$ values lie within the
$(1-\alpha{})$ quantiles of the t--distribution (denoted by
$\tquantilsgl$ in the following). This results in a quadratic
equation, the solution of which gives us three cases: (a) If the
denominator $\meanx$ is significantly different from zero at
significance level $\alpha$ (that is, if $\meanx^2/\vxx > \tquantil$),
we obtain a normal confidence interval with the limits $l_{1}$ and
$l_{2}$ (``bounded case''):
\begin{equation}
\label{eq:fieller:limits:orig}
  l_{1/2} = 
  \frac{(\meanx~\meany - \tquantil\vxy)\pm
    \sqrt{(\meanx~\meany-\tquantil\vxy)^2 - 
         (\meanx^2-\tquantil\vxx)(\meany^2-\tquantil\vyy)}}
  {\meanx^2-\tquantil\vxx}
\end{equation}
If the denominator $\meanx$ is not significantly different from zero,
we first need to calculate:
\begin{eqnarray}
  \tunbounded   & = & \frac{\meanx^2}{\vxx} + \frac{(\meany\vxx{} - 
              \meanx\vxy{})^2}{\vxx{}(\vxx{}\vyy{} - \vxy{}^2)}
\end{eqnarray}
With this we can discriminate between:  (b) If $\tunbounded > \tquantil$,
we obtain a confidence set which excludes only the values between
$l_1$ and $l_2$, but all other values are included
(``unbounded/exclusive'' case). (c) If $\tunbounded < \tquantil$, the
confidence set does not exclude any value at all (``unbounded'' case).

This might seem as quite a complex behavior, but it is possible to
present these results in a simple, geometrical fashion which is
equivalent to Fieller's method
(\citeNP{vonLuxburg_Franz_04_TR133,Guiard_89}; see also
\citeNP{Milliken_82}).  For this, we depict $X$ at the abscissa and
$Y$ at the ordinate of a coordinate system and draw a line from the
origin to the estimates $(\meanx,\meany)$; as is shown in
Figure~\ref{fig:fieller:fullexample}a.  The slope of this line
corresponds to the ratio ($\frac{\meany}{\meanx}$) and is graphically
represented by the intersection of the line with a vertical line at
$X=1$. Now we need to determine the confidence limits for the ratio.  

\inserthere{Figure~\ref{fig:fieller:fullexample}}

Because all points which lie inside the gray wedge project onto the
same interval, all we need to do is to adjust the size of this wedge
such that the appropriate confidence level for the ratio is achieved.
\citeA{vonLuxburg_Franz_04_TR133} showed that the wedge forms tangents to
an ellipse centered at $(\meanx,\meany)$. The projection of the
ellipse onto the abscissa corresponds to the marginal confidence
interval of $\meanx$, the projection onto the ordinate corresponds to
the marginal confidence interval of $\meany$ and the shape of the
ellipse is determined by the covariance $\vxy$.

Using this geometrical method, we can assess the qualitative behavior
of the Fieller confidence limits. If the denominator $(\meanx)$ is
significantly different from zero at a significance level of $\alpha$,
then the ellipse does not touch the y--axis and we get
normal, bounded confidence intervals
(Figure~\ref{fig:fieller:fullexample}a). Now assume the denominator is
not significantly different from zero such that the ellipse touches
the y--axis. In this case the result of the projection of the wedge
onto the $X=1$ line is unbounded: We either get a confidence set which
exclude only a small part of all possible values (unbounded/exclusive
case, see Figure~\ref{fig:fieller:fullexample}b; the arrows indicate
that the confidence set is unbounded), or a confidence set which does
not exclude any value at all (unbounded case,
Figure~\ref{fig:fieller:fullexample}c).

Unbounded confidence sets are certainly a puzzling result and some
remarks are necessary here: (a) for practical applications, we usually
want bounded confidence intervals. A necessary and sufficient condition
for this is that the $(1-\alpha)$ confidence interval of the
denominator does not contain zero (which is equivalent to the
denominator being significantly different from zero at a significance
level of $\alpha$).  (b) if the denominator is not significantly
different from zero, then its confidence interval allows values
arbitrarily close to zero. In consequence, the ratio can assume
arbitrarily large (or small) values and the confidence sets are
unbounded. This implies that at the given confidence level we learned
only little from our experimental data (in the unbounded/exclusive
case), or nothing at all (in the unbounded case).  While this might be a
discomforting result, it is a simple consequence of the ratio we are
interested in and there is no way to force a different outcome. In
fact, different researchers \cite{Gleser_Hwang_87,Koschat_87,Hwang_95}
have shown that any method which is not able to generate unbounded
confidence limits for a ratio can lead to arbitrary large deviations
from the intended confidence level (which I will call the
``Gleser--Hwang theorem''). We will see that this theorem limits
almost all of the alternatives to the Fieller method, except for a
special bootstrap method (the Hwang--bootstrap method) and for the
case that the true ratio is bounded away from zero.

Note, that the unbounded confidence sets contribute to the overall
performance of the method. That is, if in a certain situation there
are on average, say $10\%$ unbounded confidence sets, these will count
as being including the true ratio. For a discussion of the conditional
confidence level, given that the Fieller confidence limits are bounded,
see \citeA{Buonaccorsi_Iyer_84}. This leads to an interesting problem:
If we assume that we report a measured ratio only if it has bounded
confidence intervals, then we effectively use the conditional
confidence level. This can, however, be arbitrarily low (this follows
from the Gleser--Hwang theorem because this conditional procedure will
never generate unbounded confidence limits; 
see also \citeNP{Neyman_54}, \citeNP{Tsao_98}, and \citeNP{Read_83}).
One solution was proposed by \citeA{Tsao_Hwang_98} who suggest to
estimate the confidence as $1$ in the unbounded case and as $1-\alpha$
in the other cases (see also \citeNP{Kiefer_77}).
\subsection*{Alternative approaches}

In this section I give an overview of alternatives to Fieller's method
as discussed in the statistical literature or employed by previous
studies. (I will not discuss Bayesian approaches here, because they
are based on a different notion of confidence limits and a full
treatment would go beyond the scope of this article.
For application of Bayesian approaches to ratios see
\citeNP{Mandallaz_Mau_81,Buonaccorsi_Gatsonis_88,Raftery_Schweder_93}).

\subsubsection*{Taylor method}

The Taylor method (sometimes also called delta--method) calculates a
linear approximation for the sample estimates:
\label{COVER5}
\begin{eqnarray}
\label{eq:taylor:simple:reformulate}
\frac{\meany}{\meanx} & \approx &
\EYEX \left( 1 - \frac{\meanx}{E(X)} + \frac{\meany}{E(Y)} \right)
\end{eqnarray}
Because the approximation is linear, it is easy to calculate
confidence limits for this function if we again assume that
$(\Pmeanx,\Pmeany)$ is bivariate normally distributed. The approximate
confidence limits (denoted by $l_{1}$ and $l_{2}$) are symmetric
relative to the sample estimates $(\meanx,\meany)$ and will never be
unbounded:
\begin{eqnarray}
\label{eq:taylor:limits}
  l_{1/2} & = &
   \estEYEX \pm 
   \tquantilsgl  \left\lvert \estEYEX \right\rvert
   \sqrt{\frac{\vxx}{\meanx^2} + 
         \frac{\vyy}{\meany^2} -
         2\frac{\vxy}{\meanx \meany}} 
\end{eqnarray} 

The Taylor approximation has virtues because the linear function is
mathematically easy to handle. However, the approximation will fail
for the ``problematic'' cases, when the denominator is close to zero
(this is to be expected by the Gleser--Hwang theorem because the
Taylor limits are never unbounded). But, if the denominator has a small
\cv the Taylor method provides a serious alternative to the Fieller
method. We will see in the simulations that the Taylor method is a
very good approximation in these cases (cf.
\citeNP{Cox_90,Polsky_etal_97,Gardiner_etal_01}).

\subsubsection*{Bootstrap methods}

The bootstrap (\citeNP{Efron_79,Efron_Tibshirani_93}) is a general
purpose method which allows to determine confidence limits in an easy
and consistent way, even for very complicated statistics. It uses the
measured sample as a basis for re-sampling with the goal to create an
approximation to the population distribution.  For our ratio problem
with $N$ paired measurements, bootstrap methods would draw a large
number of samples (with replacement) from the set of the measured
values $(x_i,y_i)$. Each sample has the same size as the original
sample and we would calculate for each sample the ratio
$\frac{\meany}{\meanx}$. The distribution of these re-sampled ratios
(the ``empirical distribution'') is the basis for the calculation of
the confidence intervals. In the simplest case, the confidence
intervals are the $(1-\alpha)$ percentiles of the empirical
distribution (``percentile method''). Other methods perform certain
corrections, most notably the widely used \BCa method (``bias
corrected and accelerated''). These standard bootstrap methods can
provide an alternative to approximative solutions as the
Taylor method, especially in cases where $(\Pmeanx,\Pmeany)$ is not
normally distributed
\cite{Chaudhary_Stearns_96,Polsky_etal_97,Briggs_etal_99,Briggs_etal_02}.

However, standard bootstrap methods face two problems when dealing
with ratios: (a) Bootstrap confidence limits can be erroneous if the
variance of the statistic does not exist as in the case of ratios
\cite{Athreya_87,Knight_89}. (b) Because bootstrap confidence limits
cannot result in unbounded confidence limits the Gleser--Hwang theorem
applies such that there can be arbitrary large deviations from the
intended confidence level for ratios. We will see in the simulations
that this problem occurs if the denominator is close to zero.

\citeA{Hwang_95} showed that these problems can be overcome by a
special bootstrap method which does not perform the bootstrap on the
ratio directly, but on $T_0$ in Equation~(\ref{eq:fieller:T0}). The
method first uses the bootstrap to determine the $(1-\alpha)$
quantiles of $T_0$ and then proceeds as the Fieller method does (i.e.,
solves the quadratic equation). Depending on the result of the
quadratic equation, this method can produce unbounded confidence limits
and is therefore the only alternative to Fieller's method which is not
limited by the Gleser--Hwang theorem.  We will see in the simulations
that the Hwang--bootstrap method performs as well as the Fieller
method if the sample sizes are large enough. In addition,
\citeA{Hwang_95} showed that for non--normal distributions with
non--zero skewness, the Hwang--bootstrap method is superior to
Fieller's method:
The Fieller method is only first order correct, with the coverage
converging as $O(1/\sqrt{N})$ against the desired coverage, while the
Hwang--bootstrap method is second order correct, converging as
$O(1/N)$ (see also \citeNP{Hall_86_basic,Hall_88} for first/second
order correctness). While this qualifies the Hwang--bootstrap method
as an excellent alternative to the Fieller method, it should also be
clear that the standard bootstrap methods will not always fail. The
Hwang--bootstrap is, however, more general and it is therefore safer
to use this method than the standard bootstrap methods.

\subsubsection*{Index method}

This method can only be applied to the special case of $N$ paired
measurements. The idea is to determine the ratio for each of the $N$
subjects individually:
\begin{equation}
  r_i = \frac{y_i}{x_i}
\end{equation}
From these individual ratios (or ``indices'') the mean
$\overline{r_i}$ and standard error are calculated. Assuming that the
mean is approximately normally distributed, confidence limits are
calculated.

The index method is used very often (almost all of the example studies
in the supplementary material provided with this article used this
method).
We can justify the method in the context of a linear model if the
denominator is bounded away from zero and if the data have a specific
heteroscedastic structure, such that the the numerator has larger
variability at larger values of the denominator. This model will be
discussed in the section ``\WhenCanWeUseIndices''. 

Because the method is used so often and because it seems unlikely that
the data in all these cases show the specific heteroscedastic
structure (the studies typically do not report having tested for
this), I will first discuss what happens if the method is applied to
bivariate normal data. We will see that in this case the method can
lead to large deviations from the desired confidence level.  Also, if
the mean ratio $\overline{r_i}$ is used as point--estimate for $\EYEX$
it shows systematic biases and can be much more variable than the
ratio of the means $\meany/\meanx$.

\subsubsection*{Zero--variance method}

Some studies (e.g.,
\citeNP{Glover_Dixon_01_JEP,Glover_Dixon_01_EBR,Glover_Dixon_02_PP,Glover_Dixon_02_EBR,Haffenden_etal_01,Danckert_etal_02})
estimated the variability of the ratio by dividing the individual
numerator--value of each subject by the overall mean of the
denominator (calculated across all subjects):
\begin{eqnarray}
  r_i          & := & \frac{y_i}{\meanx}
\end{eqnarray}
From these individual ratios the mean and standard error were
calculated. It is easy to show that this procedure is equivalent to
dividing the mean of the numerator and its standard error by $\meanx$,
such that we get as estimates for the mean ratio $\estEYEX =
\frac{\meany}{\meanx}$ and its standard error: $\srr =
\frac{\smeany}{\meanx}$. An inspection of the formulas shows that this
procedure does not take into account the variability of the
denominator.  Clearly, this is problematic.  To justify this approach,
we would have to assume that the measured denominator corresponded to
the true value of the denominator in the population such that the
variability of the denominator were zero (for this reason, I call this
approach the zero--variance approach). In consequence, the
zero--variance approach will often underestimate the variability of
the ratio and will therefore result in liberal statistical tests.

\subsection*{A numerical example}

Before investigating the methods systematically, I will give a simple
example. The example is taken from a study by \citeA{PangGW02}, the
only study I found to provide enough detail to calculate the
confidence limits using most of the different methods (because the
sample sizes were small, the bootstrap would not make sense here and
is omitted). The study reported four different ratios (which I denote
with ``P1'', ``P2'', ``P3'', ``P4'') using the index
method. For more details about the data see the supplementary material
provided with this article. Also, as a tutorial example, raw data and
results are given for one of the ratios in
Table~\ref{tab:pang-02-example}, such that it is possible to
reconstruct the calculations.

\inserthere{Table~\ref{tab:pang-02-example} \& Figure~\ref{fig:pang-02-graph}}

Figure~\ref{fig:pang-02-graph}~a.--d. shows the $95\%$ confidence
limits according to each of the methods. There are large differences:
In some cases, all alternatives to the Fieller method lead to a much
smaller width of the confidence intervals than the Fieller method.
For example, in case ``P1'' the Fieller method gives an upper limit of
$498$, while all other methods estimate upper limits below $10$.  The
discrepancy occurs because the denominator is just about significantly
different from zero. If there were slightly more variability in the
denominator, then it would not be significantly different from zero
and the Fieller limits would be unbounded. This can be seen if we use
the geometric construction method for the Fieller limits
(Figures~\ref{fig:pang-02-graph}e. and~f.). In case ``P1'' the ellipse
almost touches the y--axis (which would result in unbounded limits).

The discrepancy suggests that the coverage of the alternative methods
is smaller than intended. We will see in the simulations that this
is indeed often the case. Of course, the alternative methods are not
always as bad as in case ``P1''.  This can be seen in case ``P2''
where all methods lead to similar results.

\subsection*{Simulations}

We saw that the alternatives to Fieller's method likely can lead to a
much smaller coverage than intended.
But how general is this problem? I will present the results of Monte
Carlo simulations which used known distributions of $(X,Y)$. All
methods were applied to the simulated data, and the percentage of
simulation runs in which the confidence limits contained the true
values was determined. For a $95\%$ confidence level, we expect that
the confidence limits contain the true value in $95\%$ of the
simulation runs, while in $5\%$ the confidence limits should not
contain the true value (i.e., be significantly different from the true
value).  A liberal construction method will lead to a higher
percentage of significant results, while a conservative construction
method will lead to a lower percentage.

\subsubsection*{Methods}

Each simulation is described in terms of the sample size $N$ of the
paired measurements and the \cvs{} of numerator and denominator. For
simplicity the correlation is assumed to be zero, such that we explore
a 3--dimensional parameter--space ($\Pcvx$,
$\Pcvy$, $N$). This space is covered across typical ranges in the Figures %
\ref{fig:simulation-contourplot1-N20} and %
\ref{fig:simulation-contourplot1-N500}. %
The use of the \cvs allows us to compare the results of different
studies with the simulations (see the data points which are plotted on
top of the Figures %
\ref{fig:simulation-contourplot1-N20} and %
\ref{fig:simulation-contourplot1-N500}). %
More details on these studies are given in the supplementary material
provided with this article. Note that the simulations are based on a
correlation of zero which will not be the case in the example studies.
However, further simulations showed that the results are essentially
identical for a wide range of correlation coefficients ($-.99$ \dots
$+.99$), such that this choice is not critical.

The random number generation for the normally distributed numerator
and denominator was performed using an algorithm described by
\citeA{Kinderman_Ramage_76} as implemented in the free data analysis
language R \cite{Rcite}.  $95\%$ confidence limits were calculated
according to each of the methods, and the coverage of the true ratio
$\EYEX$ was determined.

As ``standard'' bootstrap method I used the \BCa method as described
in \citeA{Davison_Hinkley_97} and implemented in the R--boot package
(the S~original was implemented by Angelo Canty, the R--port by Brian
Ripley). Results for the percentile method were similar to the \BCa
method and are therefore not shown. The number of bootstrap
replications was always $B=2000$.  For the Hwang--bootstrap method, I
performed a \BCa bootstrap on $T_0$ in Equation~(\ref{eq:fieller:T0}).
Because the
Hwang--bootstrap method is relatively new, I describe it in more
detail here: We have a sample of paired measurements $(x_i,y_i)$ with
$i=1 \dots N$ and want to bootstrap $T_0=T_0(\meanx,\meany,\EYEX)$ in
Equation~(\ref{eq:fieller:T0}). For this, we generate $B$
bootstrap--samples. Each bootstrap--sample consists of $N$~pairs drawn
with replacement from the original sample $(x_i,y_i)$. For each
bootstrap--sample, we calculate the means $(\Bmeanx,\Bmeany)$.
Following \citeauthor{Hwang_95} (\citeyearNP{Hwang_95}, p. 163/164 and
p. 170), we use these bootstrap means to determine the empirical
distribution of $T^*_0=T_0(\Bmeanx,\Bmeany,\estEYEX)$. Based on this
empirical--distribution we determine the quantiles of $T^*_0$ and then
proceed as with the Fieller method. Note, that the Hwang--bootstrap
method (as well as the Fieller method) is not restricted to paired
measurements, but can also be adapted to the case of independent
observations \cite{Fieller_54,Hwang_95,Lee_Lin_04}. In this case, the
denominator of $T_0$ will be different to reflect the different
estimate for the standard deviation of $\meany - \EYEX \meanx$. 

For the index method, additional calculations were performed using
trimmed means and winsorized standard deviations as described by
\citeA{Tukey_McLaughlin_63}. Trimming was always $25\%$ (cf.
\citeNP{Rosenberger_Grasko_83}).

\subsubsection*{Results \& Discussion}

Figure~\ref{fig:simulation-contourplot1-N20} shows the results of the
simulations for small sample sizes ($N=20$). The empirical confidence
levels of the Fieller method are not shown because they are always
close to $95\%$. The Hwang--bootstrap method performs equally well,
while most other methods are only accurate if the \cv of the
denominator is small. The zero--variance method fails if the \cv of
the denominator is larger than that of the numerator. This leads to
large deviations from the desired confidence level even in cases where
the denominator has a small \cv. Therefore we should not use the
zero--variance method. 

\inserthere{Figure~\ref{fig:simulation-contourplot1-N20}}

In Figure~\ref{fig:simulation-contourplot1-N20} all other methods are
accurate if the denominator is typically significantly different from
zero (this is left of the solid, vertical line). We might be tempted
to infer from this that as soon as the denominator is significantly
different from zero, all these methods are accurate. However, this is
not the case for the index method, as can be seen if we increase the
sample size.

Figure~\ref{fig:simulation-contourplot1-N500} shows the results for
larger sample sizes ($N=500$). The area where the denominator is
typically significant (again: left of the vertical solid line) is now
larger and stretches further to the right.  Accordingly, the area
where the bootstrap (\BCa) and Taylor methods are accurate also
stretches further to the right.

\inserthere{Figure~\ref{fig:simulation-contourplot1-N500}}

For the index method, however, this area moved to the
left. That is, by increasing the sample size we loose accuracy in this
method. This problem occurs right in the area where most of the
example studies are located (as indicated by the points plotted on top
of the Figures~\ref{fig:simulation-contourplot1-N20}
and~\ref{fig:simulation-contourplot1-N500}). A closer look shows that
there is a band of correct confidence levels for denominator \cvs of
about $1$. We will see that left of this band the index
method usually overestimates the value of $\EYEX$ and right of this
band the index method underestimates it.  These biases
lead to the deviations from the desired confidence level.

This can be seen in Figure~\ref{fig:sim-graph} which shows the results
of 40~simulations as error--bar plots. The simulations were performed
left of the band (point ``A''), in the band (point ``B''), and right
of the band (point ``C''). We expect about $2$~significant deviations
from the true ratio (given $95\%$ confidence limits and
$40$~simulation runs) and this is what we find for Fieller method,
both bootstrap methods, and the Taylor method. This is not surprising
because we are for all these methods in unproblematic areas where the
denominator is typically significant (because the results were
similar, only the results for the Fieller method are shown;
significant deviations are denoted by exclamation marks in the lower
part of the figure).

\inserthere{Figure~\ref{fig:sim-graph}}

For the index method, however, we can see two things: (a) The
results are much more variable than with the other methods.  This is
due to the fact that the index method uses the
individual ratios $\frac{y_i}{x_i}$ as the basis for calculating the
estimator. For some of these individual ratios, the denominator will
``hit'' the problematic region around zero and this will lead to huge
deviations in either positive or negative direction.  The other
methods do not have this problem because they first reduce the
variability of the denominator by calculating its mean.  (b) There is
first a tendency to overestimate the ratio $\EYEX$ (point ``A''), then
the estimate is noisy but balanced (point ``B''), and finally there is
a systematic underestimation (point ``C'').

Note, that the biases cannot be eliminated by using trimmed means and
winsorized standard deviations. Trimming excludes systematically a
certain percentage of the most extreme values from the statistics
\cite{Tukey_McLaughlin_63,Dixon_Tukey_68,Rosenberger_Grasko_83}.
Further simulations showed that trimming does indeed reduce the huge
variability of the point estimator, but does not reduce the bias and
therefore does not lead to better confidence limits, as can be seen in
the corresponding panels in
Figures~\ref{fig:simulation-contourplot1-N20} %
and~\ref{fig:simulation-contourplot1-N500}. %

In summary, the index method can fail badly if applied to bivariate
normal data even in situations in which the denominator is
significantly different from zero. In these situations, the Taylor and
standard bootstrap methods both perform gracefully, while they fail if
the denominator is not significantly different from zero. The
zero--variance method fails if the denominator has larger variability
than the numerator, while Hwang--bootstrap and Fieller methods never
fail.

\subsection*{Recommendations for the standard case}

Based on the previous discussion we can issue the following
recommendations for the ``standard'' case that the means of numerator
and denominator are approximately normally distributed (cf.
Table~\ref{tab:standard}): Fieller method and Hwang--bootstrap method
can generally be used, with the Hwang--bootstrap method having
advantages if there are deviations from normality.  If the denominator
is clearly significantly different from zero ($\Pcvmeanx < 1/3$ if
$95\%$ confidence limits are intended), we can also use Taylor and the
standard bootstrap methods. Note, that with sample sizes smaller than
$N=15$ the bootstrap methods (including the Hwang--bootstrap) lead to
slightly smaller coverage than intended (for sake of brevity these
simulations are not shown).

\inserthere{Table~\ref{tab:standard}}

Index method and zero--variance method are problematic and should not
be used. This does not mean that studies which used the index or
zero--variance methods necessarily need to be wrong. For both method
there are areas in the Figures~\ref{fig:simulation-contourplot1-N20}
and~\ref{fig:simulation-contourplot1-N500} where these methods lead to
the intended confidence level. For the index methods this is the case
if the \cv of the denominator is so small that the \emph{individual}
denominator values will hardly ever get close to zero (for $95\%$
confidence limits this corresponds to $\Pcvx < 1$ and $\Pcvx < 0.03$
for $N=20$ and $N=500$, respectively). For the zero--variance method
this is the case if the \cv of the numerator exceeds the \cv of the
denominator. Also note that the index method can be appropriate if the
data show a specific form of heteroscedasticity, see the section
``\WhenCanWeUseIndices''.

\section*{\WhenCanWeUseRegressionMethods}

We can view a ratio as the slope of a linear relationship with zero
intercept. Therefore the question arises whether we could use standard
regression methods to estimate the ratio and its confidence limits ---
which would be easier and more flexible than the methods discussed so
far. And indeed, this is sometimes possible. However, we have to be
careful about the assumptions we make. We will see that the most
critical question is whether there is error in the measurement of the
regressor (corresponding to the denominator of the ratio). Depending
on this, we might have to use the more complex measurement error
models instead of standard regression models. In the first part of
this section I will describe under which conditions we can choose
regression models and in the second part I will describe situations
which can be dealt with by regression methods.

\subsubsection*{Measurement error models vs. standard regression}

To give an overview, I will describe the linear model in a form which
allows for measurement error in the response as well as in the
regressor (cf.
\citeNP{Madansky_59,Fuller_87,Schaalje_Butts_93,Buonaccorsi_94,Buonaccorsi_95,Cheng_VanNess_99}).
Consider a regression model on true values:
\begin{equation}
\label{eq:linear:model}
  v_i = \alpha + \beta u_i + e_i
\end{equation}
with $(u_i,v_i)$ being the true values of the paired measurements
$(x_i,y_i)$.  The error $e_i$ is often called ``error in the
equation'' and assumed to be i.i.d. with zero mean and constant
variance. To specify the model, we also need to know whether the $u_i$
are random (the structural case) or whether they are fixed (the
functional case, cf.  \citeNP{Kendall_51,Kendall_52,Dolby_76}).

Often, there is measurement error (or ``error in the variables''),
such that the observed values $(x_i,y_i)$ do not correspond to the
true values $(u_i,v_i)$. Typically, the errors are assumed to be
additive:
\begin{eqnarray}
\label{eq:measurement:error}
  x_i & = & u_i + c_i \\
  y_i & = & v_i + d_i \nonumber
\end{eqnarray}
with the measurement errors $c_i$ and $d_i$ each assumed to be i.i.d.
with expected values zero and all being uncorrelated with the $u_i$
and the $e_i$ of Equation~(\ref{eq:linear:model}). Using this model we
can discuss the standard regression model, as well as measurement
error models.

First, assume that the true values can be observed exactly such that
$x_i=u_i$ and $y_i=v_i$ (i.e., $c_i$ and $d_i$ are zero and with zero
variance). This results in the model:
\begin{equation}
\label{eq:model:classic:regression}
  y_i = \alpha + \beta x_i + e_i
\end{equation}
This is the classic regression situation and we can use standard
regression methods in both, the structural as well as the functional
cases (e.g, \citeNP{Madansky_59,Sampson_74}).  

Second, assume that there is measurement error in the response $v_i$
(i.e., $d_i$ has non--zero variance), while we still can observe the
regressor exactly such that $x_i=u_i$. This results in the model:
\begin{equation}
\label{eq:model:response:error}
  y_i = \alpha + \beta x_i + e_i + d_i
\end{equation}
This model is similar to the classic regression model of
Equation~(\ref{eq:model:classic:regression}) and indeed we can use
standard regression methods if we assume that $d_i$ has constant
variance.
However, the estimation of the parameters can be improved if we have
information about the measurement error, which is typically obtained
by repeated measures on the same subject. With this information, it is
also possible to account for non--equal variances of $d_i$
\cite{Buonaccorsi_94,Buonaccorsi_95}, as well as for non--additive
errors \cite{Buonaccorsi_89,Buonaccorsi_96}.

Third, assume that the true values cannot be observed exactly (i.e.,
$c_i$ and $d_i$ have non--zero variance). In this case we have to use
measurement error models. There is a large variety of different
models. Standard textbooks are \citeA{Fuller_87} and
\citeA{Cheng_VanNess_99}.
Here, I will sketch two variants: (a) a ``classic'' structural
errors--in--variables model. This model is interesting because it
shows the typical issues related to measurement error models as well
as the connection to the Fieller method. (b) the ``Berkson model of a
control experiment'', which offers an alternative solution if we have
control over $x_i$ and which allows us to use standard regression
methods.

For the ``classic'' structural errors--in--variables model assume that
there is no ``error in equation'' (i.e., $e_i$ is zero with zero
variance) and that the measurement errors $c_i$ are uncorrelated with
the measurement errors $d_i$.  Also, assume that the true values $u_i$
and the measurement errors are normally distributed.  These
assumptions create a bivariate normal distribution for the pair of
observable variables $(x_i,y_i)$.
The model has three important properties: (i) If we ignored the
measurement error and used a standard regression procedure, this would
lead to a downward bias in the estimate for the slope $\beta$. This
bias is often called ``attenuation'' or ``regression dilution'' (cf.
\citeNP{Spearman_1904,Schmidt_Hunter_96,DeShon_98,Frost_Thompson_00,Charles_05}).
The importance of this issue can be seen by the fact that attenuation
was a key argument in a recent discussion on semantic priming in
Psychology
\cite{Greenwald_etal_96,Draine_Greenwald_98,Dosher_98,Klauer_etal_98,Miller_00,Klauer_Greenwald_00}.
Note, however, that attenuation is only a problem if we estimate
$\alpha$ or $\beta$. If we only want to predict $y$ given a certain
$x$ then we can use standard regression procedures. Also, if the
measurement errors are correlated it is possible that there is not
attenuation but that the slope is overestimated by standard regression
procedures \cite{Schaalje_Butts_93}.  (ii) The model is
nonidentifiable as long as we don't have additional information about
the error--variances such that we cannot obtain a unique solution (cf.
\citeNP{Reiersol_50,Madansky_59}). This additional information can,
for example, be the ratio of the variances of the measurement errors
$c_i$ and $d_i$ which could be estimated by repeated measure methods.
(iii) If the intercept is zero, the nonidentifiability problem
disappears and the appropriate solution is the Fieller method. This
shows the connection between measurement error models and Fieller
method.

In the ``Berkson model of a control experiment'' we assume that we
have control over $x_i$, even though we cannot measure the
corresponding true values accurately. This enables us to observe
$y_i$ at fixed, predefined $x_i$--values. If we assume that the
measurement errors $c_i$ have zero mean, then we get a model which is
quite different from the classic measurement error model: While in the
classic measurement error model the true values $u_i$ and the
measurement errors $c_i$ are uncorrelated, now the $u_i$ and the $c_i$
are perfectly negatively correlated.  \citeA{Berkson_50} showed that
in this case we can use standard regression methods (see also
\citeNP{Madansky_59,Fuller_87,Cheng_VanNess_99}). For a discussion of
this model in the context of repeated measure designs with multiple
subjects, see \citeA{Buonaccorsi_Lin_02}.

In summary, we can use standard regression methods if:
(a) we can measure the $x_i$ very accurately.
(b) we do not estimate $\alpha$ or $\beta$, but only want to
predict $y$, given a certain $x$. %
(c) we have control over $x_i$, even though we cannot measure the
corresponding true value accurately (``Berkson model of a
controlled experiment'').

\subsubsection*{Application of regression methods to ratios}

With regression models the situation is simpler and we can apply
standard methods as are described in most textbooks.  This easily
allows us to estimate $\alpha$, $\beta$ and the corresponding
confidence intervals. 
If we assume the intercept $\alpha$ to be zero, then $\beta$
corresponds to our ratio of interest.
Regression methods can, if applicable, also help us with more
complicated situations. For example, if we want to compare two or more
ratios obtained in different groups we can use the analysis of
covariance (ANCOVA). For this we set up the model:
\begin{equation}
\label{eq:ancova:compare:ratios}
  y_{gi} = \beta_g x_{gi} + e_{gi}
\end{equation}
with $x_{gi}$, $y_{gi}$ being the values obtained in group $g=1...m$
for participant $i=1...n$; $\beta_g$ the ratios of interest and
$e_{gi}$ the errors. Standard ANCOVA methods then allow to decide
whether the ratios are different \cite{Miller_86}. 

Note, however, that even in situations in which we can use regression
methods we sometimes need Fieller's method.  This is the case if we
want to calculate ratios of the parameters estimated by regression
methods, as for example in the inverse prediction described in the
Introduction. Another classic example is the slope ratio assay
\cite{Finney_78}. Here, researchers first calculate an ANCOVA as
described in Equation~(\ref{eq:ancova:compare:ratios}), but then are
interested in the ratio of two $\beta_g$ estimates (typically
indicating the effectiveness of a drug relative to a standard drug).
Again, they need Fieller's method for this ratio of regression
parameters. In general, it is possible to calculate Fieller confidence
limits for linear combinations of parameters of general linear models
\cite{Zerbe_78}, generalized linear models \cite{Cox_90}, and
mixed--effects models \cite{Young_etal_97}. For examples of such
applications see \citeA{Buonaccorsi_Iyer_84} and \citeA{Sykes_00}.

Ratios of estimated parameters can also occur in the context of
nonlinear regression models (e.g., \citeNP{Bates_Watts_88}) and
nonlinear mixed effects models (e.g.,
\citeNP{Davidian_Giltinan_95,Vonesh_Chinchilli_97,Pinheiro_Bates_02}).
These models are designed to deal with general nonlinear problems and
therefore can also deal with ratios. In addition, the nonlinear mixed
effects models can handle repeated measure data, as are frequent in
psychological and biological research.  Typically, these models
perform a linear approximation at the point of the estimated
parameters and therefore can fail in a similar way as the Taylor
method discussed in this article. But, if the denominator of the ratios
have small \cvs, these models will provide an elegant solution such
that it can be beneficial to reformulate a statistical problem in
terms of a nonlinear model (see also \citeNP{Cox_90}).
\section*{\WhenCanWeUseIndices}

Linear models as described in the previous section assume that the
residual errors are constant over the range of observations
(``homoscedastic''). Sometimes this is clearly not the case (the
errors are ``heteroscedastic''). Two classes of models can improve
this situation and lead to an interest in indices. Both models use
standard regression methods, such that as soon as the models are
specified the specification of confidence limits for the ratio of
interest pose no additional problems.  For simplicity of presentation,
I will assume in the following that the denominator of the ratios is
bounded away from zero such that it cannot attain values close to
zero. This is often the case in situations in which indices are used
(cf.  \citeNP{Belsley_72}). Without this assumption we cannot justify
the use of indices. Also, I assume that the regressor can be measured
with negligible error, such that we don't need to use measurement
error models.

First, consider the case that we want to fit a structural
regression model to our data: 
\begin{equation}
\label{eq:simple:additive:model:proportional}
y_i=\alpha + \beta x_i + e_{1,i}
\end{equation}
Assume that the residual errors $e_{1,i}$ are heteroscedastic. This
can lead to serious deviations from the desired confidence level if we
used standard regression methods to determine confidence limits for
$\alpha$ and $\beta$. Often it is possible to correct for the
heteroscedasticity by using weighted least squares analysis (e.g.,
\citeNP{Miller_86}). Sometimes, it turns out that the errors are
proportional to the absolute size of $x_i$ such that the $y_i$ spread
out with larger $x_i$. In this case, we can use a special variant of
weighted least--squares analysis and divide the whole equation by
$x_i$ \cite{Kuh_Meyer_55,Firebaugh_Gibbs_85,Kronmal_93}:
\begin{equation}
\label{eq:simple:additive:model:proportional:divided}
\frac{y_i}{x_i}=\alpha \frac{1}{x_i} + \beta + e_{2,i}
\end{equation}
If now the assumptions of regression models are met, most notably
that the new error term $e_{2,i}=e_{1,i}/x_i$ is homoscedastic and
normally distributed, then we can determine confidence limits using
standard regression methods (with $y_i/x_i$ and $1/x_i$ being response
and regressor, respectively). Note, that although this method uses
indices, it estimates the same parameters $\alpha$ and $\beta$
as the standard linear model in
Equation~(\ref{eq:simple:additive:model:proportional})
\cite{Firebaugh_Gibbs_85}.
It is also possible to have more than one regressor; an example is
given in the next section
(``\BewareSpuriousCorrelationsAndFaultyRatioStandards'') in
Equation~(\ref{eq:partial:regression:model2:divided}).  These models
are often used in econometrics, where the ratios are often called
``deflated variables'' and the denominator ``deflator''. For a
discussion of non--random denominators see \citeA{Belsley_72} and for
a discussion of the case with measurement error in the denominator see
\citeA{Casson_73}.

Using this model we can also justify the index method, if we assume
that $\alpha$ in
Equation~(\ref{eq:simple:additive:model:proportional:divided}) is zero:
\begin{eqnarray}
\label{eq:simple:index:model}
  \frac{y_i}{x_i} & = & \beta + e_i
\end{eqnarray}
with $\beta$ being the ratio of interest and $e_i$ being the error,
typically assumed to be i.i.d. as normal. Note the specific
heteroscedastic structure we have to assume to justify this method.

Second, consider an allometric or power function model
\cite{Kleiber_1947,Sholl_1948,Nevill_etal_92,Nevill_Holder_94,Nevill_Holder_95_modeling,Dreyer_Puzio_01}:
\begin{equation}
\label{eq:simple:allometric:model}
y_i=\beta \medspace x_i^\gamma \medspace e_i
\end{equation}
With $(x_i,y_i)$ being the observed values, $\beta$ and $\gamma$ the
parameters, and $e_i$ the error term.  Sometimes the parameters can be
estimated by log--transformation to a log--linear model. This results
in:
\begin{equation}
\label{eq:simple:log:linear:model}
log(y_i)=log(\beta) + \gamma log(x_i) + log(e_i)
\end{equation}
If the assumptions of regression models are met for the log-linear
model, most notably that $log(e_i)$ is homoscedastic and normally
distributed, then we can use standard regression methods on
Equation~(\ref{eq:simple:log:linear:model}) to determine the
confidence limits of $log(\beta)$ and $\gamma$.

Allometric models can also incorporate more than two variables. A good
example is given by \citeA{Nevill_etal_92} who showed that for
recreational runners the 5--km run speed ($z_i$) is well predicted by
an index of maximum oxygen uptake ($y_i$) and body mass ($x_i$). For
this, they used the allometric model:
\begin{equation}
\label{eq:allometric:runners}
z_i=\beta \medspace y_i^{\gamma_1} \medspace x_i^{\gamma_2} \medspace e_i
\end{equation}
and fitted it with log--linear regression. As result, they obtained
the fit:
\begin{equation}
\label{eq:allometric:runners:results}
z_i = 84.3 \medspace \frac{y_i^{1.01}}{x_i^{1.03}}
\end{equation}
Because the exponents are close to one, the fit contains essentially
an index, such that in this case the use of an index seems warranted.
Also, the model turned out to be superior to linear models and it
seems biologically plausible that performance is affected by oxygen
uptake relative to body mass. (For a further discussion of allometric
models and the relation to indices see also
\citeNP{Nevill_Holder_95_modeling,Nevill_Holder_95_revisited,Kronmal_93,Kronmal_95}).

In summary, there are two models which make accepted use of index
variables: The linear model with correction for heteroscedasticity by
division and the allometric model. Both models assume that the
denominator is bounded away from zero and heteroscedastic structures
of the data, with the $y_i$ spreading out in a fan--like fashion with
larger $x_i$ values.

\section*{\BewareSpuriousCorrelationsAndFaultyRatioStandards}
Spurious correlations are a famous and much discussed problem (e.g.,
\citeNP{Pearson_1897,Kronmal_93,McShane_95,Nevill_Holder_95_revisited,Kronmal_95}).
We will see that spurious correlations can occur if numerator and
denominator of a ratio are linearly related with non--zero intercept
and inappropriate methods are used, typically involving indices.

Consider that we are interested in the relationship between two
measurements $y_i$ and $z_i$, but want to ``correct'' for the effect
of a third variable $x_i$. A famous, hypothetical example was given by
\citeA{Neyman_52}: A researcher relates the number of babies to the
number of storks in a number of different counties.  Because larger
counties inhabit more women and consequently more babies (and more
storks), the researcher wants to correct for the number of women.
Table~\ref{tab:storks} gives a simplified version of the data.

\inserthere{Table~\ref{tab:storks}}

First, consider the accepted way to do the correction: For this we use
partial regression analysis and set up a restricted model and a full
model:
\begin{eqnarray}
\label{eq:partial:regression:model1}
        y_i & = & \alpha_{rest} + \beta_{rest} x_i + e_{1,i} \\
\label{eq:partial:regression:model2}
        y_i & = & \alpha_{full} + \beta_{full} x_i + \gamma z_i + e_{2,i}
\end{eqnarray}
With $y_i$ being the number of babies, $x_i$ the number of women, and
$z_i$ the number of storks; $x_i$ and $z_i$ assumed to be random and
measured with negligible error. We can think of partial regression as
a two--step process: We first fit the restricted
model~(\ref{eq:partial:regression:model1}). This model is designed to
linearly predict the babies ($y_i$) based on the number of women
($x_i$). In the second step, we determine how much the fit is improved
if we use the full model~(\ref{eq:partial:regression:model2}) which
also includes the number of storks ($z_i$). (If $z$ were a factor, we
would replace $\gamma z_i$ by parameters indicating the effects at the
different factor levels. This would correspond to an ANCOVA; cf.
\citeNP{Maxwell_etal_85}).  Visual inspection of the example data in
Table~\ref{tab:storks} shows that the number of babies is almost
perfectly predicted by the number of women, such that the addition of
the storks does not improve the fit significantly. Therefore, we
conclude that number of storks has no influence on the number of
babies.

This standard partial regression analysis assumes again that the
errors $e_{1,i}$ and $e_{2,i}$ are homoscedastic. If the errors are
heteroscedastic and scale with the size of $x_i$, we can apply the
correction discussed in the section ``\WhenCanWeUseIndices'' and
divide both equations by $x_i$. This results in:
\begin{eqnarray}
\label{eq:partial:regression:model1:divided}
        \frac{y_i}{x_i} & = & \alpha_{rest} \frac{1}{x_i} + \beta_{rest} + e_{3,i} \\
\label{eq:partial:regression:model2:divided}
        \frac{y_i}{x_i} & = & \alpha_{full} \frac{1}{x_i} + \beta_{full} + \gamma \frac{z_i}{x_i} + e_{4,i}
\end{eqnarray}
If now the errors are homoscedastic, we can proceed as before with the
standard regression methods. Note, that these corrected models
estimate the same parameters $\alpha_{rest}$, $\alpha_{full}$, $\beta_{rest}$, and
$\beta_{full}$ as the models~(\ref{eq:partial:regression:model1}) and
(\ref{eq:partial:regression:model2}) \cite{Firebaugh_Gibbs_85}.

Now, consider the problematic way to do the correction: Here, we
simply divide the number of babies ($y_i$) and storks ($z_i$) by the
number of women ($x_i$) and then investigate the linear relationship
between these individual ratios. This results in:
\begin{eqnarray}
\label{eq:index:variables}
        \frac{y_i}{x_i} & = & \beta_{full} + \gamma \frac{z_i}{x_i} + e_{5,i}
\end{eqnarray}
Typically, $\gamma$ is tested against zero. This corresponds to a
comparison of the full model~(\ref{eq:index:variables}) with the
restricted model:
\begin{eqnarray}
\label{eq:index:variables:restricted}
        \frac{y_i}{x_i} & = & \beta_{rest} + e_{6,i}
\end{eqnarray}
Comparing these models to the partial regression models with
correction for heteroscedasticity in the
Equations~(\ref{eq:partial:regression:model1:divided})
and~(\ref{eq:partial:regression:model2:divided}) shows that the models
are equivalent if we assume that $\alpha_{rest}$ and $\alpha_{full}$
in Equations~(\ref{eq:partial:regression:model1:divided})
and~(\ref{eq:partial:regression:model2:divided}) are zero. Because
$\alpha_{rest}$ and $\alpha_{full}$ correspond to the intercepts in
Equations~(\ref{eq:partial:regression:model1})
and~(\ref{eq:partial:regression:model2}) this means that we assume the
intercepts of the linear relationships between $y_i$ and $x_i$ to be
zero. This assumption of zero intercept is at the core of the debate
about spurious correlations \cite{Kuh_Meyer_55,Firebaugh_88}. The
problem is that if the intercepts deviate from zero then the
correction does not work properly.

In our stork example, this can be seen in Table~\ref{tab:storks}: The
birth--rate $y_i/x_i$ is highly and significantly correlated with the
stork--rate $z_i/x_i$, such that based on this problematic analysis we
would conclude that there is a strong dependence. This dependence,
however, is only ``spurious'' and is generated by the fact that the
intercepts in the Equations~(\ref{eq:partial:regression:model1})
and~(\ref{eq:partial:regression:model2}) are not zero (if we
extrapolate the data, there are $y_0 \approx 10$ babies at $x_0=0$
women).

Now, one might argue that there is indeed theoretical reason to assume
that the intercept should be zero: Obviously, if there are no women,
there cannot be any babies (\citeNP{McShane_95}, but see:
\citeNP{Kronmal_95}). However, this zero--point can easily be obtained
if the data are non--linear beyond the range of observations. An
every--day example would be the fuel--efficiency of cars.  For longer
distances, the fuel consumed is linear to the distance traveled. For
short distances, however, this linearity breaks down because here cars
need an disproportionate large amount of fuel. Therefore, we would be
wrong if we compared the fuel efficiency (expressed as a ratio: miles
per gallon or liters per 100~kilometers) of one car that was used for
short distances with that of another car that was used for long
distances.

Of course, the assumption of zero intercept is not always wrong. But,
given all the potential problems involved if it is violated it should
be carefully tested or there should be serious theoretical reasons to
assume a linear model with zero intercept.  There are ample examples
of studies which likely fell prey to spurious correlations (cf.
\citeNP{Kuh_Meyer_55,Kronmal_93}). Also note that the problematic
method described above relies on a second strong assumption, namely
the assumption of heteroscedasticity with the errors scaling with the
size of $x_i$.  This should also be tested.  A good example of a
careful model--test which also considered the potential
heteroscedasticity of the data is the study of \citeA{Nevill_etal_92}
on the ratio of maximum oxygen uptake and body mass in recreational
runners (see the section ``\WhenCanWeUseIndices'').

A problem closely related to spurious correlations is the problems of
faulty ratio standards. If ratios are used to define a medical
standard for the ``normal'' or average human, and if the data have
non--zero intercept, then this standard can lead to serious biases.
For example, \citeA{Tanner_1949} showed that stroke volume of the
heart is linearly related to body weight with positive, non--zero
intercept. Because, however, the average \emph{ratio} was used as
standard, the average lightweight person was automatically above the
standard and the average heavy person was automatically below the
standard.  \citeA{Tanner_1949} gives a large number of further
illustrative examples and concludes that many patients classified as
having deviant values ``may have been suffering from no more
formidable a disease than statistical artefact'' (p. 3).

In summary, spurious correlations and faulty ratio standards can occur
if we use ratios on data which are linearly related, but with
non--zero intercept. In the literature, this problem is typically
discussed together with the use of indices, but it is not restricted
to indices. The use of indices only adds the additional assumption of
heteroscedasticity with the errors scaling with the size of the
denominator.
\section*{Summary and Conclusions}

Ratios of measured quantities pose unusual statistical problems. When
dealing with ratios we should first clarify whether we are justified
in using a ratio, that is whether the numerator can safely be assumed
to be a linear function of the denominator with zero intercept.
Otherwise we should better use a linear model with non--zero
intercept; see the section
``\BewareSpuriousCorrelationsAndFaultyRatioStandards''.  If a ratio is
appropriate, we can use the methods summarized in
Table~\ref{tab:standard} and describe in the section
``\TheStandardCase''. 

Sometimes, we can simplify the calculations by using standard
regression methods. This is typically the case if the denominator can
be measured with negligible error; see the section
``\WhenCanWeUseRegressionMethods''. Regression methods can help us
also in more complicated cases in which, for example, we want to
compare ratios or have repeated measure data. If it turns out that the
residuals are heteroscedastic and if the denominator is bounded away
from zero, index method or allometric models can be potential
remedies, see the section ``\WhenCanWeUseIndices''.

This shows that the wide use of the index method (illustrated by the
example studies provided with this article) rests on very specific and
likely often problematic assumptions. The simulations in the section
``\TheStandardCase'' show that the index method can lead to large
deviations from the desired confidence level if these assumptions are
not met. Also, the point estimate closely associated with this method
(i.e., the mean ratio) can lead to systematic biases and much more
variable estimates than the ratio of the means. Therefore we should
not use the index method as long as there is no indication for the
specific heteroscedastic structure assumed by this method. A simple,
straightforward alternative for cases in which one might be tempted to
use the index method (i.e., if the denominator is bounded away from
zero) is the Taylor method. For this, all the the researcher needs to
do is to use Equation~(\ref{eq:taylor:limits}) instead of the index
method. Of course, all the other methods described in
Table~\ref{tab:standard} would also be viable alternatives (most
notably, standard bootstrap methods if there are deviations from
normality). If, however, the denominator is not bounded away from
zero, we best use either the Fieller method or the Hwang--bootstrap
method.

\newpage 
\bibliography{lit-general,lit-vf,lit-ratio-users}

\newpage
\section*{Author note} 

\noindent Volker H. Franz, University of Giessen, Giessen, Germany. \\
\vspace{1cm}

\noindent I wish to thank Ulrike von Luxburg and Brian White for 
helpful comments on this manuscript. 

\vspace{1cm}

\noindent Correspondence should be addressed to: \\
\noindent Mail address: \\
Volker Franz, \\
Justus--Liebig--Universität Giessen  \\
FB 06 / Abt. Allgemeine Psychologie  \\
Otto--Behaghel--Strasse 10F \\
35394 Giessen, Germany \\
\noindent \parbox[b]{1.5cm}{Phone:} ++49 (0)641 99--26112\\
\noindent \parbox[b]{1.5cm}{Fax:}   ++49 (0)641 99--26119 \\
\noindent \parbox[b]{1.5cm}{Email:} volker.franz@psychol.uni-giessen.de \\

\newpage
\section*{Figure Legends}
\def\labelenumi{\bf{Figure \arabic{enumi}:} }
\begin{enumerate}
  
\item \label{fig:fieller:fullexample} Qualitative behavior and
  geometric construction of the Fieller confidence limits for
  $\EYEX$. The confidence limits (indicated by the thick,
  solid vertical lines) can be constructed using a wedge which forms
  tangents to an ellipse centered at $(\meanx,\meany)$. The size of this
  ellipse is such that its projection onto the abscissa corresponds to
  the marginal confidence interval of $E(X)$, the projection onto the
  ordinate corresponds to the marginal confidence interval of $E(Y)$
  and the shape of the ellipse is determined by the covariance $\vxy$.
  If the denominator is significantly different from zero at a
  significance level of $\alpha$, then the ellipse will not touch the
  ordinate and the $(1-\alpha)$ confidence limits will be bounded (left
  panel). If the denominator is not significantly different from zero,
  the ellipse will touch the ordinate and the confidence limits will
  be unbounded (middle and right panel, the arrows indicate infinity).
  In the unbounded/exclusive case, we still can exclude a small interval
  (the dashed vertical line in the middle panel) while in the unbounded
  case, we can not exclude any value at all (right panel).
  
\item \label{fig:pang-02-graph} {\bf a.--d.} Comparison of the $95\%$
  confidence limits for the example data from \citeA{PangGW02}, as
  calculated by the Fieller, Taylor, index and zero--variance methods.
  The study reported four different ratios in four different
  conditions (denoted here by ``P1'', ``P2'', ``P3'', ``P4''). For P1,
  the upper Fieller limit is 498 (which is beyond the upper limit of
  the y--scale). {\bf e.--f.}  The geometrical construction method
  applied to the conditions P1 and P2.  At P1 the construction ellipse
  just about touches the y--axis, which leads to an almost infinite
  upper Fieller limit. If the ellipse touched the y--axis, we would
  get unbounded/exclusive Fieller limits.
  
\item \label{fig:simulation-contourplot1-N20} Empiric confidence
  levels of the different methods for small sample sizes ($N=20$). The
  empiric confidence levels of the Fieller method are not shown,
  because they are of course always close to the expected $95\%$.  The
  empiric confidence levels are color--coded. For example, light gray
  corresponds to an empiric confidence level between $90\%$ and
  $99\%$. Left of the solid vertical line the denominator is typically
  significantly different from zero. 
  (This is achieved by depicting
  $\Pcvmeanx=\frac{\Psdmeanx}{E(X)}=0.5$ which corresponds to
  $\Pcvx=\frac{\Psdx}{E(X)}=2.2$ at the abscissa). The dotted diagonal
  line indicates equal \cvs{} of numerator and denominator. In each
  panel the \cvs that were reported in a number of example studies are
  plotted as single data points (see also the supplementary material
  provided with this article). At the \cvs{} indicated with ``A'',
  ``B'', ``C'', and ``D'' further simulations were run, see
  Figure~\ref{fig:sim-graph}.
  
\item \label{fig:simulation-contourplot1-N500} Empiric confidence
  levels for $N=500$. As in
  Figure~\ref{fig:simulation-contourplot1-N20}, the denominator is
  typically significantly different from zero left of the solid
  vertical line. (This is achieved by depicting
  $\Pcvmeanx=\frac{\Psdmeanx}{E(X)}=0.5$ which corresponds to
  $\Pcvx=\frac{\Psdx}{E(X)}=11.2$ at the abscissa). For further
  detail, see Figure~\ref{fig:simulation-contourplot1-N20}.
  
\item \label{fig:sim-graph} Results of simulations at the points ``A''
  ``B'' and ``C'' (as are shown in the Figures
  \ref{fig:simulation-contourplot1-N20} and %
  \ref{fig:simulation-contourplot1-N500}). Each plot shows the results
  of $40$~simulation--runs, ordered by the magnitude of the estimated
  ratio ($\frac{\meany}{\meanx}$). In each run $N=500$ subjects are
  simulated and analyzed using the Fieller and index methods (the
  results of Hwang--bootstrap, Taylor, and bootstrap \BCa were
  practically identical to the results of the Fieller method and are
  therefore not shown). For $95\%$ confidence limits, we expect that
  about $2$ of the $40$ simulation runs are significantly different
  from the true value (as indicated by the exclamation marks in the
  bottom of each plot). For the points ``B'' and ``C'' the rightmost
  panel shows the results of the index method at a larger scale.
  Despite this larger scale, the estimated ratio of one simulation was
  still beyond the scope of the scale. This values is indicated
  numerically ($42$).  The results at point ``D'' are similar to the
  results at point ``C'' and are therefore not shown.

\end{enumerate}
\def\boxit{} 
\newpage 
\centerline{\boxit{\includegraphics[scale=0.9]{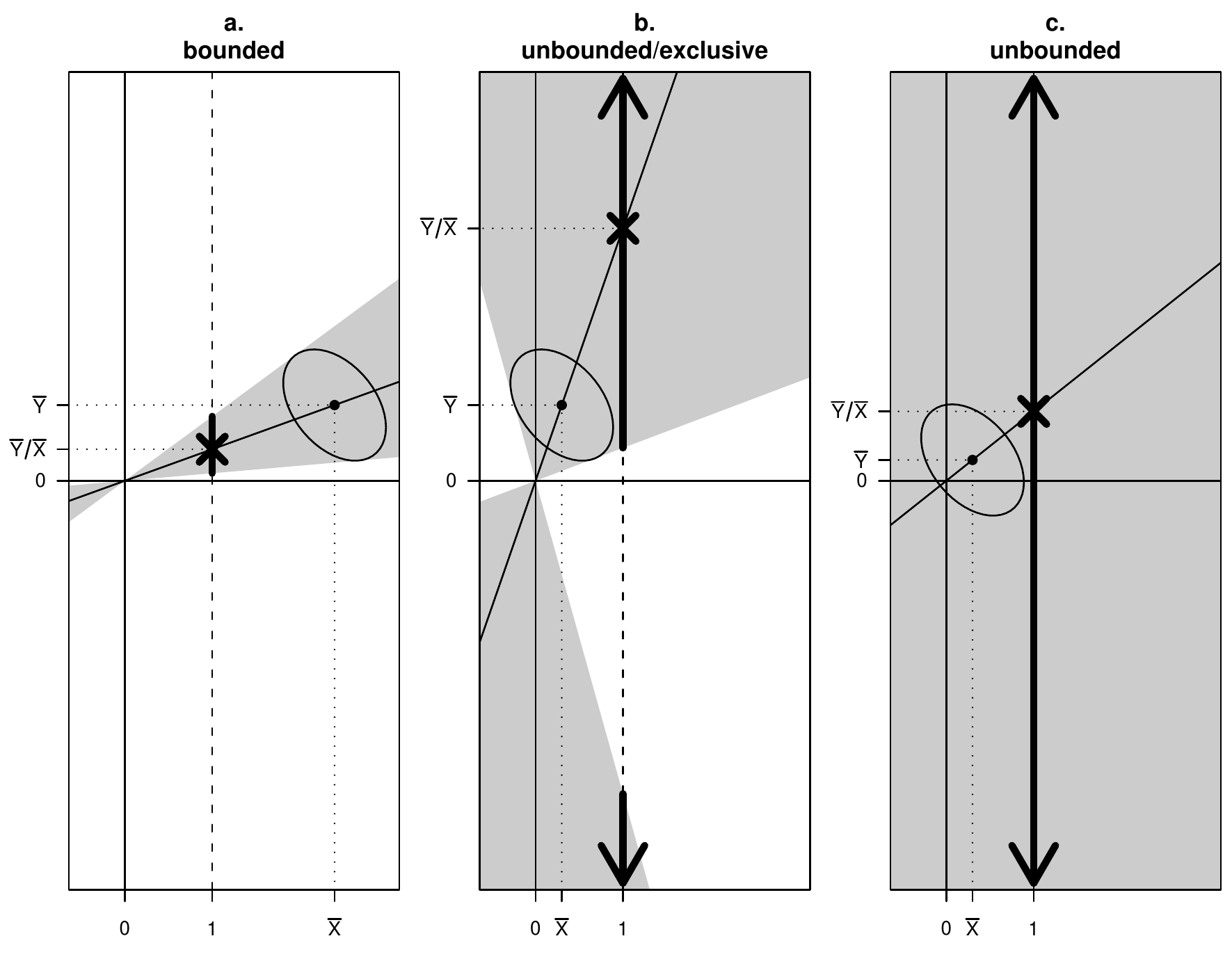}}}
\vfill{\tiny Figure~\ref{fig:fieller:fullexample}}

\newpage 
\centerline{\boxit{\includegraphics[angle=90,scale=0.9]{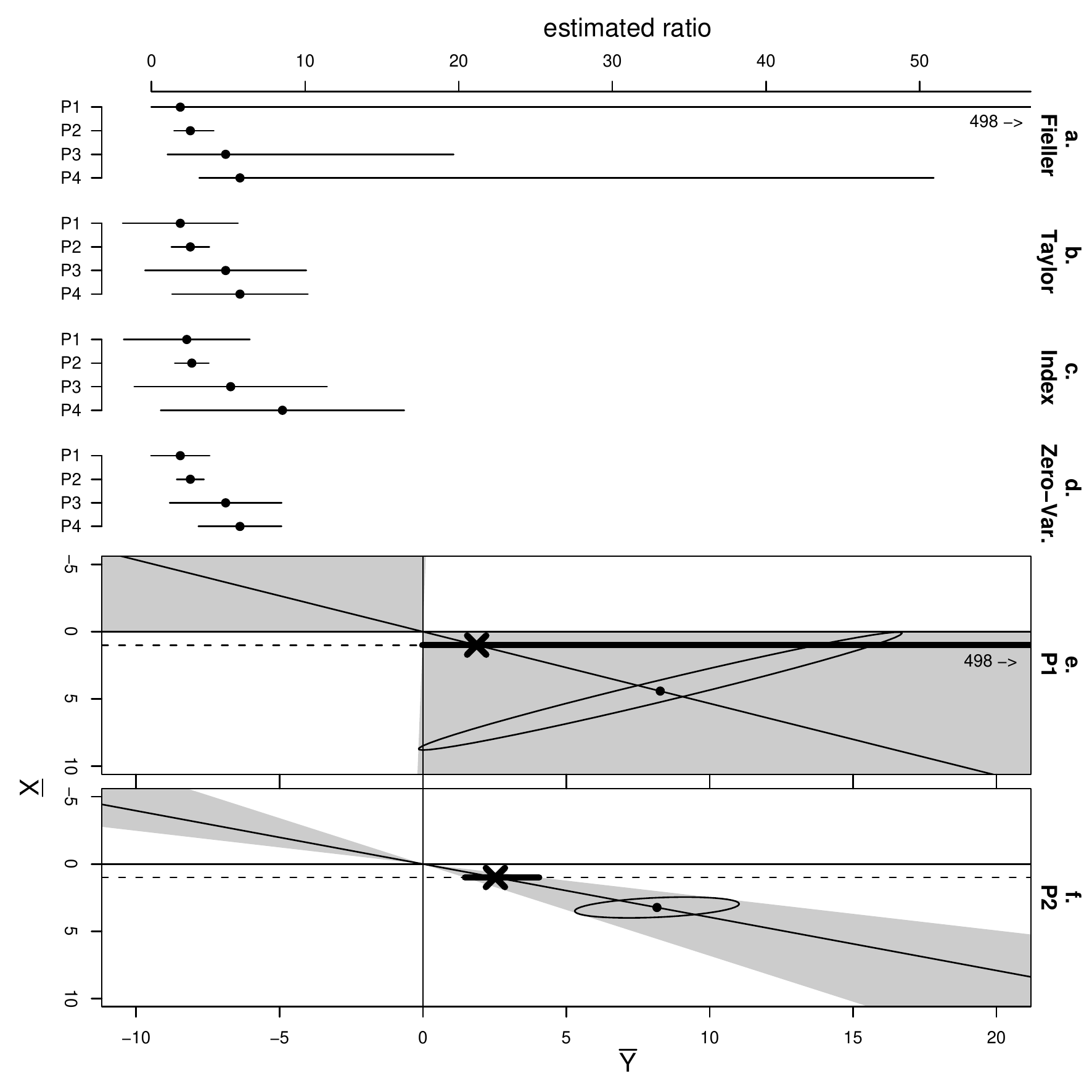}}} %
\vfill{\tiny Figure~\ref{fig:pang-02-graph}} 

\newpage 
\centerline{\boxit{\includegraphics[scale=0.8]{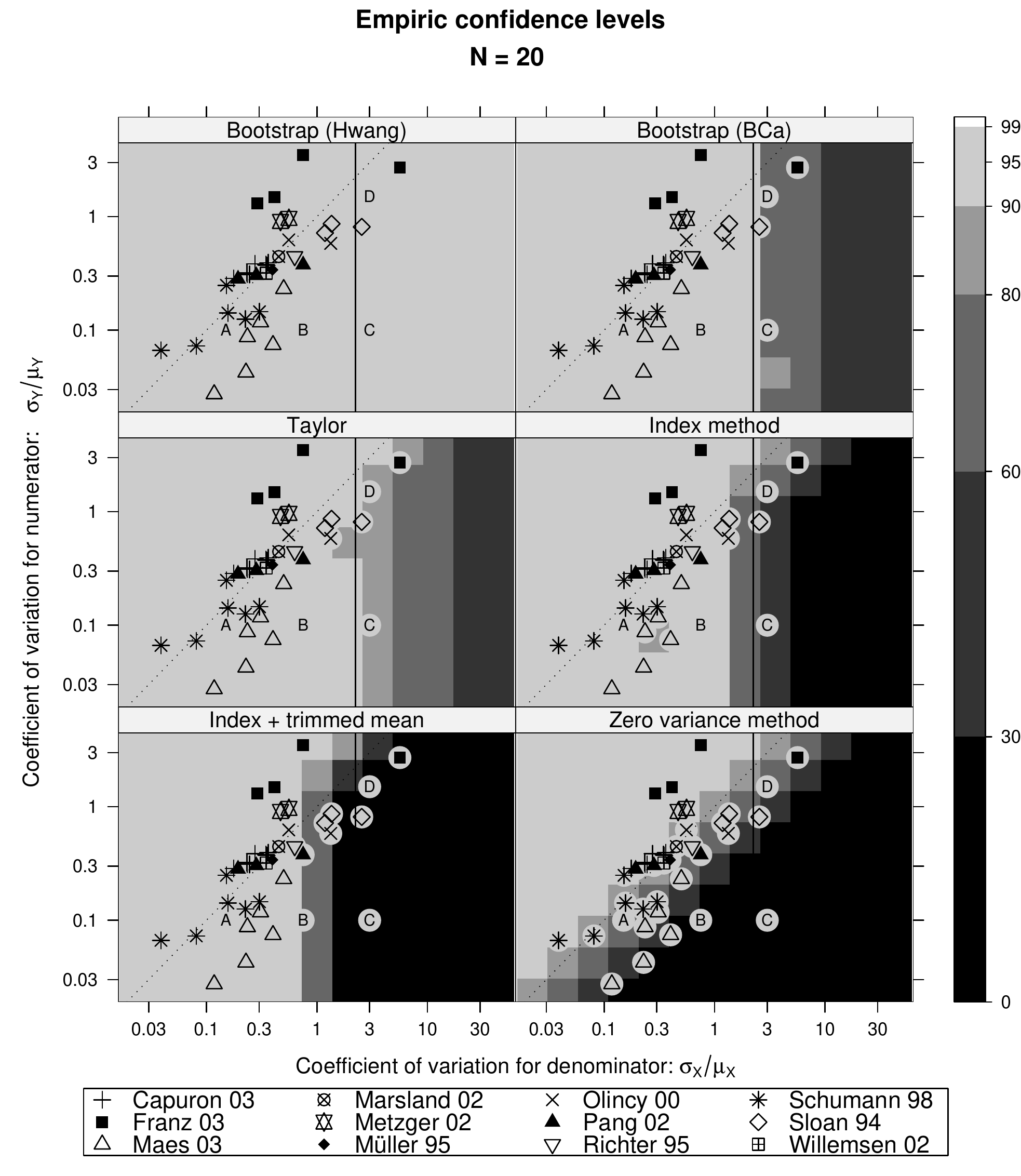}}} 
\vfill{\tiny Figure~\ref{fig:simulation-contourplot1-N20}}

\newpage 
\centerline{\boxit{\includegraphics[scale=0.8]{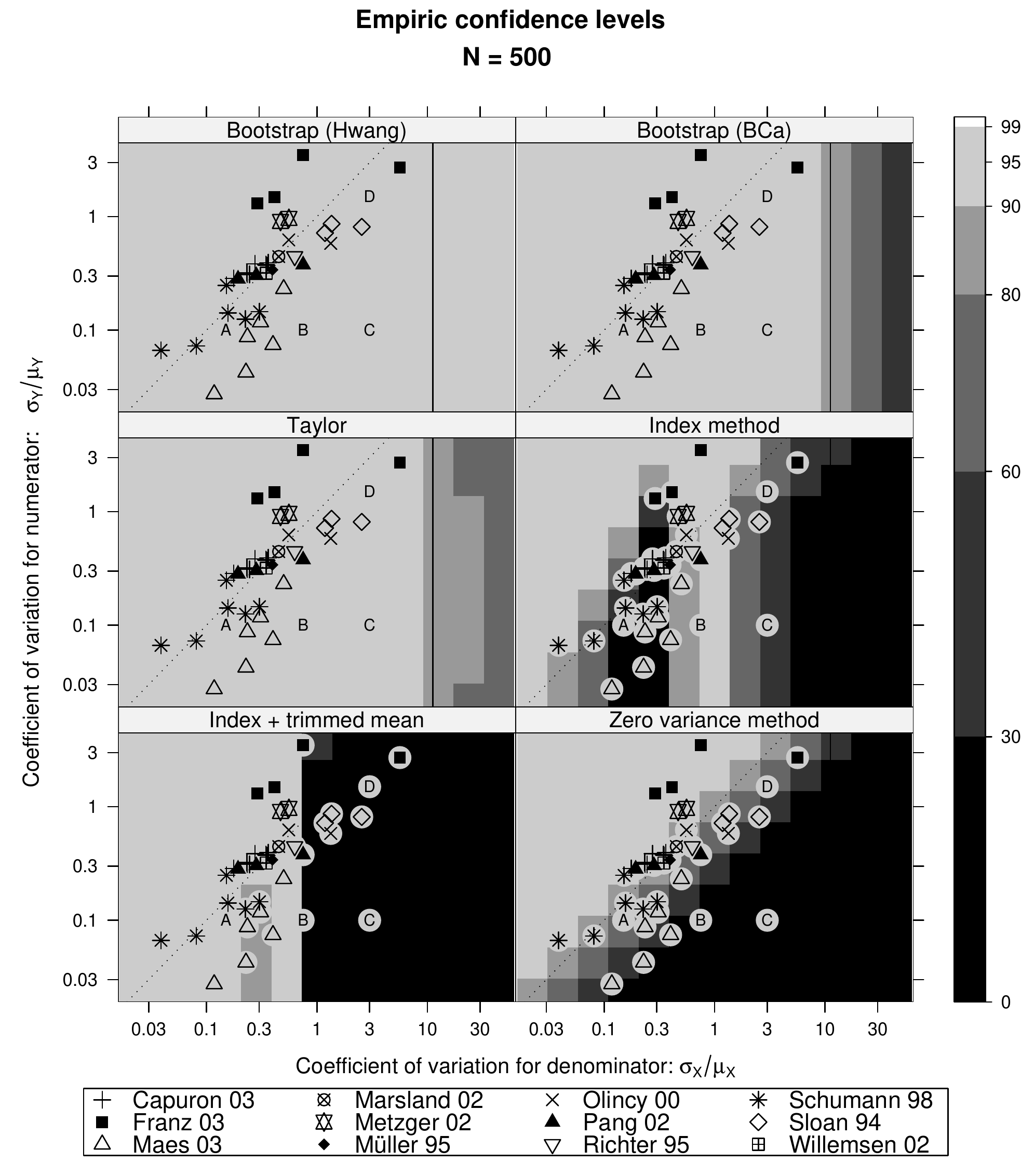}}}
\vfill{\tiny Figure~\ref{fig:simulation-contourplot1-N500}}

\newpage 
\centerline{\boxit{\includegraphics[scale=0.8]{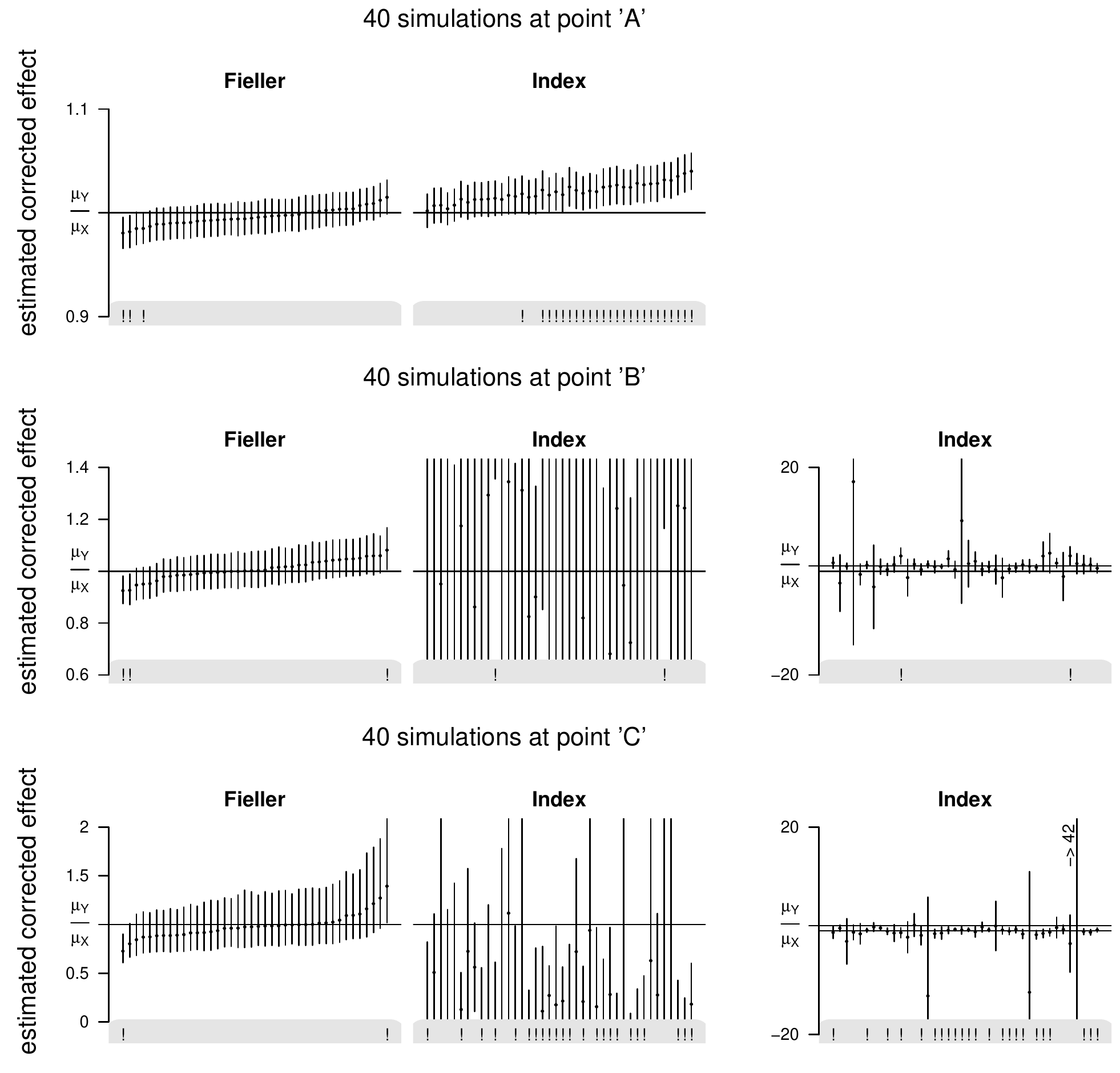}}}
\vfill{\tiny Figure~\ref{fig:sim-graph}}

\newpage
\changespacing{1}
\begin{table}[htbp]
\caption{Results of the example calculation for point ``P1'' in Figure~\protect\ref{fig:pang-02-graph}}
\begin{center}
\begin{tabular}{cccc}
  method              &  lower limit & estimate & upper limit \\
\hline
  Fieller             &  -0.02       &  1.88    & 498.75 \\
  Taylor              &  -1.88       &  1.88    &   5.64 \\
  Index               &  -1.81       &  2.29    &   6.39 \\
  Zero--variance      &  -0.03       &  1.88    &   3.79
\end{tabular}
\end{center}
\label{tab:pang-02-example}
  \par
  {\bf Note.} 
  \protect\citeA{PangGW02} 
  measured the following pairs of values:  $y_i=(4.87, 8.30,11.66)$ and 
   $x_i=(6.34,4.02,2.88)$. Calculated are $95\%$
  confidence limits, based on the quantiles of the Student--t--distribution
  $\tquantilsgl(df=2)=\pm4.3027$. For further details see the supplementary material 
  provided with this article. 
\end{table}

\newpage
\changespacing{1}
\begin{table}[htbp]
\caption{Standard methods to calculate confidence limits for ratios}
\begin{center}
\begin{tabular}{lll} 
  distribution of $(X,Y)$ & further restrictions                & adequate method \\
  \hline
  bivariate normal        &                                     & Fieller \\
  bivariate normal        & $\Pcvmeanx < 1/3$                   & Taylor \\
  not necessarily normal  & $N         > 15$                    & Hwang--bootstrap \\
  not necessarily normal  & $N         > 15$; $\Pcvmeanx < 1/3$ & Standard bootstrap \\
  \hline
\end{tabular}
\end{center}
\label{tab:standard}
  \par
  {\bf Note.} 
  The restrictions are meant as rules of thumb and apply to the
  case that we are interested in $95\%$ confidence limits. 
\end{table}

\newpage
\changespacing{1}
\begin{table}[htbp]
\caption{Hypothetical example for spurious correlations}
\begin{center}
\begin{tabular}{llllll}
  county & women ($x_i$) & babies ($y_i$) & storks ($z_i$) & birth--rate ($\frac{y_i}{x_i}$) & stork--rate ($\frac{z_i}{x_i}$)  \\
  \hline                                                        
  1      & 1             & 15.8           & 3.2            &    15.8                         &    3.2      \\
  2      & 2             & 20.2           & 4.1            &    10.1                         &    2.1      \\
  3      & 3             & 25.4           & 5.6            &     8.5                         &    1.9      \\
  4      & 4             & 30.1           & 6.3            &     7.5                         &    1.6      \\
  \hline
\end{tabular}
\end{center}
\label{tab:storks}
  \par
  {\bf Note.} 
  Women are x10000. 
\end{table}

\newpage
\section*{Supplementary material:  Details about the cited studies}

\noindent This supplement gives a short description of the studies for
which the \cvs are shown in the Figures %
\ref{fig:simulation-contourplot1-N20} and %
\ref{fig:simulation-contourplot1-N500}. %
The exact values for numerator and denominator variability and the
sample sizes are also shown in
Table~\ref{tab:data-ratio-studies-table}. If specified, I describe the
method which was used to calculate confidence limits (or SEM, which
are typically interpreted as $68\%$ confidence limits).

\paragraph*{\citeNP{CapuronNMLNFM03}:}
Data are from Table~1, p.~909. The data describe the ratio of
Kynurenine (KYN) to Tryptophan (TRP) during interferon (IFN)--$\alpha$
therapy: KYN/TRP. \emph{Method used:} Index method (p.~908)
\emph{Background:} TRP degradation into KYN by the enzyme,
indoleamine-2,3-dioxygenase, during immune activation may contribute
to development of depressive symptoms during IFN--$\alpha$ therapy.
26~patients with malignant melanoma had received IFN--$\alpha$
treatment and received in parallel either an antidepressant
(paroxetine) or placebo. \emph{Conditions:} Antidepressant free
patients vs.  paroxetine--treated patients, measured at treatment
initiation, weeks~2, 4, and 12.

\paragraph*{\citeNP{Franz_03_SV}:} 
Data are from Table~2, p. 219 and from my own records. The data
describe the effect of an illusionary change of object--size relative
to the effect of a physical change of object--size at different times
of a reach to grasp movement. The ratio
(illusion--effect)/(physical--size--effect) was calculated for each
time--point. The study is also discussed in
\citeA{Franz_04_BBS_Glover} and
\citeA{Franz_etal_05_JEP}, where also re-calculations
using the correct Fieller--method are given. \emph{Method used:}
Index method. \emph{Background:} The planning/control model of
action \cite{Glover_Dixon_01_JEP,Glover_04_BBS,Glover_02_TICS} assumes
that grasping is sensitive to certain illusionary changes of
object--size only in early stages of the movement (planning), but not
in later stages (control). In consequence, the relative effects of
these illusions should decrement during a grasping movement.
\emph{Conditions:} Grasp aperture measured at start of movement
($t=0\%$), at the time of the maximum grip apterture ($t=100\%$), and
at intermediate times ($t=25\%$, $t=50\%$, $t=75\%$).

\paragraph*{\citeNP{MaesCBLN00}:}
Data are from Table~1, p.~912. The data describe the
$\omega6$/$\omega3$ polyunsaturated fatty acids (PUFAs) ratios.
\emph{Method used:} Index method on $z$--transformed scores (p.~912).
\emph{Background:} Psychological stress in humans induces the
production of proinflammatory cytokines. An imbalance of $\omega6$ to
$\omega3$ PUFAs in the peripheral blood causes an overproduction of
proinflammatory cytokines.
The study examined whether an imbalance in $\omega6$ to $\omega3$
PUFAs in human blood predicts a greater production of proinflammatory
cytokines in response to psychological stress. \emph{Conditions:}
$\omega6$/$\omega3$ ratios a few weeks before (PRE) and after (POST)
as well as one day before (STRESS) a difficult oral examination.
Participants were also divided into groups with low/high fatty acid
status.

\paragraph*{\citeNP{MarslandHCB02}:}
Data are from Table~1, p.~867. Data describe the ratio of T--helper
(CD4+) cells to T--suppressor/cytotoxic (CD8+) cells: CD4+/CD8+.
\emph{Method used:} Not described. \emph{Background:} To explore the
stability of immune reactivity in humans, the study assessed
lymphocyte responses to a speech task and a mental arithmetic task.
Dependent measure was (beside others) the ratio of CD4+ to CD8+ cells.
\emph{Conditions:} Mental arithmetic task, Speech task, Baseline
performances.

\paragraph*{\citeNP{MetzgerCPLPPO02}:}
Data are from Table~2, p.~54. Data describe the amplitude of P50
event--related brain potentials in response to two auditory clicks
(i.e., second click amplitude/first click amplitude) \emph{Method
  used:} Index method. \emph{Background:} Individuals
with post traumatic stress disorder (PTSD) have been found to show
several event--related brain potential abnormalities including P50
suppression.  Female Vietnam nurse veterans with and without current
PTSD completed P50 paired--click tasks: Two clicks were presented and
the amplitude of the P50 for each of the clicks was determined.
\emph{Conditions:} Current PTSD versus never PTSD.

\paragraph*{\citeNP{MullerRBEH95}:}
Data are from Table~1, p.~74. Data describe the relationship of low
density lipoprotein (LDL) cholesterol to high density lipoprotein
(HDL) cholesterol: LDL/HDL. \emph{Method used:} Not described. 
\emph{Background:} The relationship between habitual anger coping
styles, especially anger expression in a socially assertive manner and
serum lipid concentrations was assessed.  The LDL/HDL ratio was
analyzed because it provides a predictor for coronary heart disease.
\emph{Conditions:} Two groups: male versus female.

\paragraph*{\citeNP{OlincyRHYMCMSAF00}:}
Data are from Table~2, p.~972. Data describe the amplitude of P50
event--related brain potentials in response to two auditory clicks
(i.e., test P50 amplitude/conditioning P50 amplitude). \emph{Method
  used:} Index method (p.~972) \emph{Background:}
Attention-deficit/hyperactivity disorder (ADHD) and schizophrenia are
both conceptualized as disorders of attention.  Failure to inhibit the
P50 auditory event--evoked response, extensively studied in
schizophrenia, could also occur in ADHD patients, if these two
illnesses have common underlying neurobiological substrates. The study
examined the inhibition of the P50 auditory event--evoked potential in
unmedicated adults with ADHD, schizophrenic outpatients, and normal
control subjects. Auditory stimuli were presented in a paired
stimulus, conditioning--testing paradigm. The ratio of the test to the
conditioning response amplitudes were observed.  \emph{Conditions:}
Three groups: unmedicated adults with ADHD, schizophrenic outpatients,
and normal control subjects.

\paragraph*{\citeNP{PangGW02}:}
Because this study gave excellent details about the data, I could use
it as an example in this article. For simplicity (and because we are
only interested in the statistical properties of the data), I used the
following aliases for the conditions:
\begin{center}
  \begin{tabular}[c]{llll}
    alias      & cell & parameter                            & source \\
    \hline
    P1         & OFF  & $\Delta_{g_{CL}}/\Delta_{g_{C}}(NR)$ & Table~1, p.~24 \\
    P2         & ON   & $\Delta_{g_{CL}}/\Delta_{g_{C}}(NR)$ & Table~1, p.~24 \\
    P3         & OFF  & $Q_C(P+S+I)/Q_C(NR)$                 & Table~2, p.~25 \\
    P4         & ON   & $Q_C(P+S+I)/Q_C(NR)$                 & Table~2, p.~25 \\
  \end{tabular}
\end{center}
Also, for simplicity, I used the absolute values in the cases P3 and
P4 (values are negative in the original study. Using the absolute
values does not change anything for our analysis). \emph{Background:}
\citeA{PangGW02} investigated the relative contributions of bipolar
and amacrine cell input to light responses of 3~and 5~retinal ganglion
cells. Two of the ratios describe the light--evoked changes in
chloride conductance relative to the cation conductance
$\Delta_{g_{CL}}/\Delta_{g_{C}}(NR)$ in normal Ringer's (NR) solution,
the other two ratios describe the light--evoked charge transfer in
picrotoxin + strychnine + Imidazole-4-acidic acid (P+S+I) relative to
NR: $Q_C(P+S+I)/Q_C(NR)$.  \emph{Method used:} Index method. Note: In
the calculation of the variances, \citeauthor{PangGW02} divided
sometimes by $N$, (describing the sample variability) and sometimes by
$N-1$ (estimating the population variability).  For consistency, I
always used $N-1$ in my calculations.  \emph{Background:} Light-evoked
postsynaptic currents (lePSCs) were recorded from ON, OFF and ON--OFF
ganglion cells in dark--adapted salamander retinal slices under
voltage clamp conditions, and the cell morphology was examined using
Lucifer yellow fluorescence with confocal microscopy. The charge
transfer of lePSCs in NR and in P+S+I was compared.

\paragraph*{\citeNP{RichterHMS95}:}
Data are from Table~1, p.~133/134. Data represent salivary [K+]/[Na+]
ratios.  \emph{Method used:} Index method with log--transformed data
(p.~137). \emph{Background:} It was hypothesized that choice
reaction--time testing would cause salivary [K+]/[Na+] to increase.
Relative contributions of [K+] and [Na+] to ratio changes were
investigated in 23 hypertensives and 10 hospital staff. Changes in
post--rest and post--test ionic concentrations and [K+]/[Na+] were
investigated.  \emph{Conditions:} 5~conditions: day~1 (relaxed), day~2
(pre--test), unpaced RT task, paced RT task, post--test (rest); Two
groups: Hypertensives and control group.

\paragraph*{\citeNP{SchumannTYMBM98}:}
Data are from Table~2, p.~1374. They represent the ratio of cerebral
blood flow (CBF) to cerebral blood volume (CBV), as measured by PET:
CBF/CBV.  \emph{Method used:} Index method and non--parametric
tests (p.~1371).  \emph{Background:} Local cerebral perfusion pressure
(CPP), a crucial parameter that should allow a better assessment of
the haemodynamic compromise in cerebrovascular diseases, is not
currently measurable by non--invasive means.  Experimental and
clinical studies have suggested that the regional ratio of cerebral
blood flow to cerebral blood volume (CBF/CBV), as measured by PET,
represents an index of local CPP in focal ischaemia.  The study was
designed to evaluate further the reliability of the CBF/CBV ratio
during manipulations of CPP by deliberately varying mean arterial
pressure (MAP) in the anesthetized baboon.  Cortical CBF, CBV,
cerebral metabolic rate for oxygen and oxygen extraction fraction were
measured by PET in 10 anesthetized baboons.  \emph{Conditions:} Five
baboons (Group A) underwent four PET examinations at different levels
of MAP: base line, moderate hypotension, minor hypotension, profound
hypotension. Five other baboons (Group B) were subjected to
hypertension and were compared with their base line state.

\paragraph*{\citeNP{SerrienW01}:}
Data are from Table~1, p.~419. Data present the ratio of grip force to
load force during grasping: grip--force/load--force. \emph{Method
  used:} Not described.  \emph{Background:} The study examined
interlimb interactions of grasping forces during a bimanual
manipulative assignment that required the execution of a
drawer--opening task with the left hand and an object--holding task
with the right hand.  The grip/load--force ratio of the bimanual task
was compared with the unimanual performance in order to investigate
the coordinative constraint between grip and load force.
\emph{Conditions:} Unimanual versus bimanual; object holding versus
drawer opening.

\paragraph*{\citeNP{SloanSBBPBSG94}:}
Data are from Table~2, p.~93. Data present the ratio of low (LF) to
high (HF) frequency bands of heart period variability (HPV): LF/HF.
\emph{Method used:} Mixed effect regression model on log--transformed
data (p.~92). \emph{Background:} The study investigated changes in cardiac
autonomic control during psychological stress in ambulatory subjects.
24--h electrocardiographic recordings of 33 healthy subjects were
analyzed for heart period variability responses associated with
periodic diary entries measuring physical position, negative affect,
and time of day. A total of 362 diary entries were made during the
24--h sessions, each in response to a device which signaled on an
average of once per hour.  HPV was analyzed in the frequency domain,
yielding estimates of spectral power in low and high frequency bands,
as well as the LF/HF ratio. \emph{Conditions:} Standing, sitting,
reclining positions.

\paragraph*{\citeNP{WillemsenCRD02}:} 
Data are from Table~1, p.~225. Data describe the ratio of T--helper
(CD4+) cells to T--suppressor/cytotoxic (CD8+) cells: CD4+/CD8+.
\emph{Method used:} Not specified. \emph{Background:} To examine
gender differences in immune reactions to stress and relationships
between immune and cardiovascular reactivity, measures of cellular and
mucosal immunity and cardiovascular activity were recorded in 77~men
and 78~women at rest and in response to active (mental arithmetic) and
passive (cold pressor) stress tasks. \emph{Conditions:} Two groups:
Men versus Women; mental arithmetic, rest, and cold pressor.

\newpage
\changespacing{1}
{\scriptsize
  \begin{longtable}{lllrrrrrrl}
    \caption{Illusion effects and corrected illusion effects of the example studies.}
    \label{tab:data-ratio-studies-table} \\
    \hline
    study/simulation & condition & N & $\meany$ & $\hat{\sigma}_y$ & $\meanx$ & $\hat{\sigma}_x$ & $\hat{\sigma}_y/\meany$ & $\hat{\sigma}_x/\meanx$ & plot \\
    \hline
    \endfirsthead
    Table~\ref{tab:data-ratio-studies-table} (continued) \\
    \hline
    study/simulation & condition & N & $\meany$ & $\hat{\sigma}_y$ & $\meanx$ & $\hat{\sigma}_x$ & $\hat{\sigma}_y/\meany$ & $\hat{\sigma}_x/\meanx$ & plot \\
    \hline
    \endhead
    \hline
    \endfoot
    \hline
    \endlastfoot
Capuron 03   & antidep-free init       & 15  & 1.6    & 0.5     & 35.9  & 8.4    & 0.312 & 0.234 & y \\ 
Capuron 03   & antidep-free week 2     & 15  & 3.7    & 1.4     & 30.4  & 10.5   & 0.378 & 0.345 & y \\ 
Capuron 03   & antidep-free week 4     & 15  & 2.8    & 1.1     & 30.7  & 11.1   & 0.393 & 0.362 & y \\ 
Capuron 03   & antidep-free week 12    & 15  & 2.8    & 0.8     & 38    & 6.7    & 0.286 & 0.176 & y \\ 
Capuron 03   & Paroxetin  init.        & 11  & 1.3    & 0.5     & 33.5  & 9.2    & 0.385 & 0.275 & y \\ 
Capuron 03   & Paroxetin week 2        & 11  & 3.5    & 1.3     & 30.8  & 10.8   & 0.371 & 0.351 & y \\ 
Capuron 03   & Paroxetin week 4        & 11  & 2.8    & 0.9     & 31.4  & 7.3    & 0.321 & 0.232 & y \\ 
Capuron 03   & Paroxetin week 12       & 11  & 2.9    & 0.9     & 32.7  & 8      & 0.310 & 0.245 & y \\ 
Franz 03     & t=0                     & 26  & 0.226  & 0.612   & 0.011 & 0.062  & 2.708 & 5.636 & y \\ 
Franz 03     & t=25                    & 26  & 0.333  & 1.16    & 0.292 & 0.219  & 3.483 & 0.750 & y \\ 
Franz 03     & t=50                    & 26  & 0.974  & 1.443   & 0.765 & 0.315  & 1.482 & 0.412 & y \\ 
Franz 03     & t=75                    & 26  & 1.278  & 1.8     & 1.04  & 0.302  & 1.408 & 0.290 & n \\ 
Franz 03     & t=100                   & 26  & 1.474  & 1.927   & 1.119 & 0.324  & 1.307 & 0.290 & y \\ 
Maes 03      & Low (PRE)               & 17  & 28.97  & 3.41    & 2.82  & 0.87   & 0.118 & 0.309 & y \\ 
Maes 03      & Low (STRESS)            & 17  & 29.85  & 2.22    & 3.15  & 1.26   & 0.074 & 0.400 & y \\ 
Maes 03      & Low (POST)              & 17  & 29.95  & 6.93    & 3.1   & 1.55   & 0.231 & 0.500 & y \\ 
Maes 03      & High (PRE)              & 10  & 33.93  & 0.93    & 5.45  & 0.64   & 0.027 & 0.117 & y \\ 
Maes 03      & High (STRESS)           & 10  & 33.25  & 2.9     & 5.67  & 1.33   & 0.087 & 0.235 & y \\ 
Maes 03      & High (POST)             & 10  & 32.92  & 1.4     & 5.49  & 1.25   & 0.043 & 0.228 & y \\ 
Marsland 02  & Arithmetic Baseline     & 31  & 705    & 314     & 388   & 175    & 0.445 & 0.451 & y \\ 
Marsland 02  & Arithmetic Task         & 31  & 699    & 302     & 393   & 173    & 0.432 & 0.440 & n \\ 
Marsland 02  & Speech Baseline         & 31  & 719    & 314     & 396   & 168    & 0.437 & 0.424 & n \\ 
Marsland 02  & Speech Task             & 31  & 736    & 314     & 449   & 210    & 0.427 & 0.468 & n \\ 
Metzger 02   & P50 current             & 24  & 1.78   & 1.72    & 4.52  & 2.52   & 0.966 & 0.558 & y \\ 
Metzger 02   & P50 never               & 24  & 1.74   & 1.58    & 4.98  & 2.33   & 0.908 & 0.468 & y \\ 
Müller 95    & Males                   & 53  & 188.85 & 34.65   & 54.98 & 12.22  & 0.183 & 0.222 & n \\ 
Müller 95    & Females                 & 33  & 115.58 & 39.34   & 57.61 & 22.78  & 0.340 & 0.395 & y \\ 
Olincy 00    & Schizophrenia           & 16  & 2.53   & 1.58    & 1.53  & 0.85   & 0.625 & 0.556 & y \\ 
Olincy 00    & ADHD                    & 16  & 2.08   & 1.21    & 0.66  & 0.88   & 0.582 & 1.333 & y \\ 
Olincy 00    & Normal                  & 16  & 2.61   & 1.57    & 0.5   & 0.65   & 0.602 & 1.300 & n \\ 
Pang 02      & P1                      & 3   & 8.277  & 3.396   & 4.413 & 1.763  & 0.410 & 0.400 & n \\ 
Pang 02      & P2                      & 5   & 8.162  & 2.31    & 3.228 & 0.623  & 0.283 & 0.193 & y \\ 
Pang 02      & P3                      & 3   & 224    & 68      & 46    & 13     & 0.304 & 0.283 & y \\ 
Pang 02      & P4                      & 5   & 278    & 105     & 48    & 36     & 0.378 & 0.750 & y \\ 
Richter 95   & Hypertensive day 1      & 23  & 18.1   & 8.8     & 6.2   & 3.3    & 0.486 & 0.532 & n \\ 
Richter 95   & Hypertensive day 2      & 23  & 25.7   & 12.1    & 6.4   & 3.7    & 0.471 & 0.578 & n \\ 
Richter 95   & Hypertensive unpaced RT & 23  & 34.7   & 15.5    & 7     & 4.4    & 0.447 & 0.629 & y \\ 
Richter 95   & Hypertensive paced RT   & 23  & 36.2   & 14.6    & 6.8   & 3.3    & 0.403 & 0.485 & n \\ 
Richter 95   & Hypertensive rest       & 23  & 29.8   & 11.4    & 6.2   & 2.8    & 0.383 & 0.452 & n \\ 
Richter 95   & Normals day 1           & 10  & 34.2   & 8.7     & 8.2   & 1.7    & 0.254 & 0.207 & n \\ 
Richter 95   & Normals day 2           & 10  & 41.7   & 11.8    & 6.5   & 1.1    & 0.283 & 0.169 & n \\ 
Richter 95   & Normals unpaced RT      & 10  & 53.8   & 23.6    & 7.5   & 2      & 0.439 & 0.267 & n \\ 
Richter 95   & Normals paced RT        & 10  & 50.7   & 21.7    & 7.3   & 2.2    & 0.428 & 0.301 & n \\ 
Richter 95   & Normals rest            & 10  & 46.4   & 12.2    & 7     & 1.5    & 0.263 & 0.214 & n \\ 
Schumann 98  & Hypotension Baseline    & 5   & 31.1   & 3.9     & 3.15  & 0.71   & 0.125 & 0.225 & y \\ 
Schumann 98  & Moderate Hypotension    & 5   & 27.5   & 4       & 3.61  & 1.09   & 0.145 & 0.302 & y \\ 
Schumann 98  & Minor Hypotension       & 5   & 24.7   & 3.5     & 3.2   & 0.5    & 0.142 & 0.156 & y \\ 
Schumann 98  & Profound Hypotension    & 5   & 19.7   & 4.9     & 3.63  & 0.55   & 0.249 & 0.152 & y \\ 
Schumann 98  & Baseline Hypertension   & 5   & 27.5   & 2       & 2.72  & 0.22   & 0.073 & 0.081 & y \\ 
Schumann 98  & Hypertension            & 5   & 36.1   & 2.4     & 2.84  & 0.11   & 0.066 & 0.039 & y \\ 
Serrien 01   & Onset Drawer U          & 6   & 20.31  & 4.52    & 13.92 & 3.34   & 0.223 & 0.240 & n \\ 
Serrien 01   & Impact Drawer U         & 6   & 24.53  & 4.04    & 15.32 & 3.18   & 0.165 & 0.208 & n \\ 
Serrien 01   & Onset Drawer B          & 6   & 21.34  & 3.84    & 14.52 & 3.5    & 0.180 & 0.241 & n \\ 
Serrien 01   & Impact Drawer B         & 6   & 23.72  & 4.35    & 15.24 & 3.25   & 0.183 & 0.213 & n \\ 
Serrien 01   & Static Object U         & 6   & 10.44  & 0.9     & 8.01  & 0.02   & 0.086 & 0.002 & n \\ 
Serrien 01   & Onset Object B          & 6   & 11.82  & 1.15    & 8.02  & 0.02   & 0.097 & 0.002 & n \\ 
Serrien 01   & Impact Object B         & 6   & 12.4   & 1.25    & 8.01  & 0.01   & 0.101 & 0.001 & n \\ 
Sloan 94     & Standing                & 96  & 1210.4 & 984.2   & 243.6 & 619.82 & 0.813 & 2.544 & y \\ 
Sloan 94     & Sitting                 & 191 & 1354.5 & 973.5   & 518.5 & 613.07 & 0.719 & 1.182 & y \\ 
Sloan 94     & Reclining               & 28  & 1392.8 & 1205.93 & 559.8 & 759.38 & 0.866 & 1.357 & y \\ 
Willemsen 02 & Men Rest                & 78  & 626    & 170     & 419   & 154    & 0.272 & 0.368 & n \\ 
Willemsen 02 & Men Artithmetik         & 78  & 589    & 143     & 423   & 154    & 0.243 & 0.364 & n \\ 
Willemsen 02 & Men Cold pressor        & 78  & 560    & 144     & 405   & 154    & 0.257 & 0.380 & n \\ 
Willemsen 02 & Women Rest              & 79  & 722    & 230     & 446   & 155    & 0.319 & 0.348 & y \\ 
Willemsen 02 & Women Artithmetik       & 79  & 694    & 203     & 473   & 184    & 0.293 & 0.389 & n \\ 
Willemsen 02 & Women Cold pressor      & 79  & 691    & 211     & 448   & 171    & 0.305 & 0.382 & n \\ 
Point A      & -                       & 500 & 1      & 0.1     & 1     & 0.15   & 0.100 & 0.150 & y \\ 
Point B      & -                       & 500 & 1      & 0.1     & 1     & 0.75   & 0.100 & 0.750 & y \\ 
Point C      & -                       & 500 & 1      & 0.1     & 1     & 3      & 0.100 & 3.000 & y \\ 
Point D      & -                       & 500 & 1      & 1.5     & 1     & 3      & 1.500 & 3.000 & y \\ 
  \end{longtable}
  \noindent
  Note:
  The column ``plot'' indicates whether the data point is plotted in the Figures %
  \ref{fig:simulation-contourplot1-N20} and %
  \ref{fig:simulation-contourplot1-N500}. %
}

\end{document}